\documentclass[sigconf]{acmart}
\usepackage{bookmark}
\usepackage{amsmath}
\usepackage{algorithmic}
\usepackage{graphicx}
\usepackage{textcomp}
\usepackage{xcolor}
\usepackage{fancyhdr}
\usepackage{hyperref}
\usepackage{xspace}
\usepackage{tabularx}
\usepackage{tikz}
\usepackage{forest}
\usepackage{cleveref}
\usetikzlibrary{positioning, shapes.geometric}
\usetikzlibrary{shadows}
\usetikzlibrary{trees}
\usepackage[most]{tcolorbox}
\usepackage{enumitem}

\newcommand{\blackcircled}[1]{%
    \tikz[baseline=(char.base)]{
        \node[shape=circle, fill=black, text=white, inner sep=0.5 pt] (char) {#1};
    }%
}

\tcbset{
  insightbox/.style={
    colback=gray!10, %
    colframe=black, %
    boxrule=0.2mm, %
    arc=1mm, %
    left=1mm, %
    right=1mm, %
    top=0mm, %
    bottom=0mm, %
    width=\linewidth, 
  }
}

\AtBeginDocument{%
  }

\copyrightyear{2025}
\acmYear{2025}
\setcopyright{cc}
\setcctype{by}
\acmConference[CCS '25]{Proceedings of the 2025 ACM SIGSAC Conference on Computer and Communications Security}{October 13--17, 2025}{Taipei, Taiwan}
\acmBooktitle{Proceedings of the 2025 ACM SIGSAC Conference on Computer and Communications Security (CCS '25), October 13--17, 2025, Taipei, Taiwan}\acmDOI{10.1145/3719027.3744816}
\acmISBN{979-8-4007-1525-9/2025/10}

\definecolor{darkviolet}{HTML}{9400D3}
\definecolor{PineGreen}{HTML}{01796F}
\definecolor{neonpink}{HTML}{FF10F0}
\definecolor{turquoise}{HTML}{53D6E0}

\newcommand{\new}[1]{{\color{black}#1}}
\newcommand{\cam}[1]{{\color{black}#1}}

\begin{document}
\sloppy
\title{A Sea of Cyber Threats: Maritime Cybersecurity from the Perspective of Mariners}

\author{Anna Raymaker}
\affiliation{
  \institution{Georgia Institute of Technology}
  \city{Atlanta}
  \state{GA}
  \country{USA}}
\email{araymaker3@gatech.edu}

\author{Akshaya Kumar}
\affiliation{
  \institution{Georgia Institute of Technology}
  \city{Atlanta}
  \state{GA}
  \country{USA}}
\email{akshayakumar@gatech.edu}

\author{Miuyin Yong Wong}
\affiliation{
  \institution{University of Maryland}
  \city{College Park}
  \state{MD}
  \country{USA}}
\email{miuyin@umd.edu}

\author{Ryan Pickren}
\affiliation{
  \institution{Georgia Institute of Technology}
  \city{Atlanta}
  \state{GA}
  \country{USA}}
\email{rpickren3@gatech.edu}

\author{Animesh Chhotaray}
\affiliation{
  \institution{Georgia Institute of Technology}
  \city{Atlanta}
  \state{GA}
  \country{USA}}
\email{achhotaray3@gatech.edu}

\author{Frank Li}
\affiliation{
  \institution{Georgia Institute of Technology}
  \city{Atlanta}
  \state{GA}
  \country{USA}}
\email{frankli@gatech.edu}

\author{Saman Zonouz}
\affiliation{
  \institution{Georgia Institute of Technology}
  \city{Atlanta}
  \state{GA}
  \country{USA}}
\email{szonouz6@gatech.edu}

\author{Raheem Beyah}
\affiliation{
  \institution{Georgia Institute of Technology}
  \city{Atlanta}
  \state{GA}
  \country{USA}}
\email{rbeyah@coe.gatech.edu}

\renewcommand{\shortauthors}{Anna Raymaker et al.}

\begin{abstract}
    Maritime systems, including ships and ports, are critical components of global infrastructure, essential for transporting over 80\% of the world’s goods and supporting internet connectivity. However, these systems face growing cybersecurity threats, as highlighted by recent attacks disrupting Maersk, one of the world’s largest shipping companies, causing widespread impacts on international trade and shipping. The unique challenges of the maritime environment—including diverse operational conditions, extensive physical access points, fragmented regulatory frameworks, and its deeply interconnected, international structure—require \new{maritime}-specific cybersecurity research. Despite the sector’s critical importance, maritime cybersecurity remains an underexplored area, leaving significant gaps in our understanding of its challenges and risks.

    To take an early step in addressing these gaps, we investigate how operators of maritime systems perceive and navigate cybersecurity challenges within the complex maritime landscape. We conducted a user study comprising surveys and semi-structured interviews with 21 officer-level mariners. Participants reported direct experiences with shipboard cyber-attacks, including offshore GPS spoofing and logistics-disrupting ransomware, demonstrating the real-world impact of these threats. Despite this, our findings reveal systemic and human-centric issues, such as cybersecurity training that is poorly designed to address the unique challenges of maritime operations, insufficient detection and response solutions, and severe gaps in mariners’  \new{understanding of cybersecurity}. Our contributions include a detailed \new{categorization} of cyber threats identified by mariners, as well as actionable recommendations for improving maritime security, including enhancements to cybersecurity training, attack response protocols, and regulatory frameworks. These insights aim to guide future research and policy to bolster the resilience of maritime systems against evolving cyber threats.

\end{abstract}

\makeatletter
\gdef\@copyrightpermission{
    \begin{minipage}{0.2\columnwidth}
        \href{ https://creativecommons.org/licenses/by/4.0/}{\includegraphics[width=0.90\textwidth]{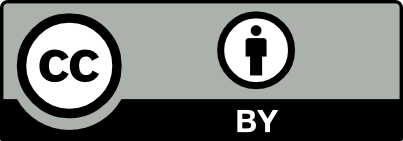}}
    \end{minipage}\hfill
    \begin{minipage}{0.8\columnwidth}
        \href{ https://creativecommons.org/licenses/by/4.0/}{This work is licensed under a Creative Commons Attribution International 4.0 License.}
    \end{minipage}
    \vspace{0pt}
}
\makeatother

\ccsdesc[500]{Security and privacy~Usability in security and privacy}

\keywords{Maritime Cybersecurity, User Study, Cyber-Physical Systems}

\maketitle

\section{Introduction}
Maritime systems, including ships, ports, and their supporting networks, are vital components of our global infrastructure. They are essential to the worldwide economy, with over 80\% of the world's goods transported by sea~\cite{unctad_maritime_transport_2024}. In addition to shipping, maritime operations play a crucial role in supporting global internet infrastructure through the construction and maintenance of undersea cables, which are increasingly at risk from sabotage~\cite{cnn2024subseacables,guardian2024underseacables}. Recent incidents underscore the importance of these maritime systems and their reliable operation. For example, in March 2024, a cargo ship lost control of its propulsion system and collided with the Francis Scott Key Bridge in Maryland, United States, killing 6~people and costing over 100 million dollars in damages~\cite{npr2024baltimore, ap2024baltimore}. Similarly, in March 2021, a cargo ship ran aground in the Suez Canal, blocking an estimated 9 billion dollars in trade a day over a six-day period~\cite{bbc2021suez, das2021suez}. 

While such catastrophic events can be due to benign failures, they can also be induced by cybersecurity attacks on maritime systems. Such threats are not just hypothetical; they have happened in reality. For example, in January 2023, a ransomware attack~\cite{greig2024ransomware} on widely used maritime software disrupted major shipping companies, including Maersk, significantly impacting global trade. In response, international maritime organizations have recently implemented cybersecurity standards that ships must comply with~\cite{dnv_maritime_cybersecurity_2024}. All these developments reflect changing winds in how this critical infrastructure sector considers cybersecurity. 

Despite the increasing focus on cybersecurity in the maritime sector, there has been limited prior research on this topic, particularly when also considering the people who interface with and operate maritime systems. The existing prior work only focuses on specific systems, attacks, or defenses (e.g., security analysis of Very Small Aperture Terminals (VSAT)~\cite{pavur2020tale}, defenses against GPS spoofing attacks~\cite{liu2021stars}), leaving systemic and human-centric issues unexplored.  In contrast,  the security of other cyber-physical systems (CPS) has received substantially more attention, including aircraft~\cite{lundberg2014security,birnbach2017wi,jansen2017localization,jansen2021trust,longo2024collision} and automobiles~\cite{rouf2010security,bhatia2021evading,checkoway2011comprehensive,cho2016fingerprinting,garcia2016lock,hu2021automated,jing2024revisiting,koscher2010experimental,tang2024eracan,wen2020plug,xue2022said,zhu2024ae}. 

Securing ships presents a distinct set of challenges compared to other cyber-physical system domains like automobiles and aircraft. Maritime environments are uniquely characterized by diverse operational conditions, extensive physical access points, and international and regional regulatory frameworks. Ships host significantly more individuals with legitimate physical access to critical systems, including third-party contractors and vendors, often numbering in the hundreds on larger vessels~\cite{britannia,maritimecyprus}. The vetting processes for these individuals frequently lack the rigor necessary to ensure security~\cite{useless_twic,twic_threat}, amplifying the risk of insider threats or inadvertent system compromise. Additionally, the maritime sector contends with varied equipment configurations, diverse crews, and port facilities with differing security standards. Mariners often work under grueling conditions, including long hours and high workloads, which can impair their ability to detect and respond to cyber-attacks. These challenges are compounded by the industry's fragmented and inconsistent application of cybersecurity standards.

These issues are further reflected in the Department of Homeland Security and United States Coast Guard’s 2024 call for public input to update maritime security regulations, emphasizing the urgent need to address these systemic vulnerabilities~\cite{us_dhs_cybersecurity_marine_2024}. Such complexities highlight the inadequacy of directly applying traditional CPS security frameworks to maritime systems. Instead, targeted research that addresses the sector’s unique vulnerabilities, operational realities, and critical infrastructure status is essential for bolstering the resilience of maritime systems against modern cybersecurity threats.

In addressing the unique challenges of securing ships, it is crucial to establish a baseline understanding of the maritime industry’s cybersecurity landscape. Unlike other domains, maritime operators are not dedicated cybersecurity experts but are still expected to manage and secure complex cyber-physical systems under diverse and often challenging conditions. No prior research has systematically explored how mariners perceive and approach cybersecurity or how factors like training and regulation shape their behaviors. By focusing on these foundational topics, we aim to provide a groundwork that future studies can build upon. This leads us to pose the following research questions:

\noindent\textbf{RQ1.} What are mariners' \new{perceptions of cybersecurity}?

\noindent\textbf{RQ2.} What are mariners' cybersecurity practices, and what role does training and regulation play in shaping these practices?

To address these research questions, we conducted a user study comprising an online survey for participant recruitment followed by semi-structured interviews. We interviewed 21 mariners holding officer-level positions (e.g., Chief Engineer, First Mate, Captain) in the shipping industry or their equivalents on military vessels. Our study revealed several surprising insights, including a disconnect between the perceived physical impact of cyber threats and their actual consequences. Additionally, two participants offered valuable perspectives from their work in sub-sea cable construction, shedding light on the cybersecurity challenges associated with protecting this essential infrastructure, which underpins global communications. These insights, alongside the broader findings, highlight the need for focused research on domain-specific cyber threats that concern mariners and the maritime industry as a whole.

\textbf{Our contributions} include a \new{categorization} of cyber threats identified by mariners, offering valuable insights into the shipping industry’s cybersecurity challenges. We also provide actionable recommendations to improve cybersecurity training, enhance cyber-attack detection and response, and develop unified and effective cybersecurity regulations. These findings aim to guide immediate improvements in maritime cybersecurity practices and inform future research to address the unique challenges of this domain.

\section{Background}
\label{sec:background}
This section provides key context about the maritime domain, \new{including the types of ships mariners operate, the roles mariners perform, the regulatory landscape they operate in, and the cybersecurity risks they face.} \new{We draw from maritime industry resources~\cite{imo_regulations,marine_insight_ship_types,britannica_ferries,houston_maritime_ship_types,lloyds_register_classification,oneocean_vessel_types,suncam_ship_types,mr_marine_ship_types,history_of_ships,career_explorer,military_sealift_command,mitags,maritime_professinal_training,dot_maritime_administration,northwest_maritime_academy,nscc,marine_agency_services,shipfinex,primonautic,anchoredemia,cslships,life_of_sailing,ship_crew_structure,woods_hole,deck_department,imo,world_shipping_council,clearseas,maritime_and_waterways,top_10_classification,heisenbergshipping,us_code,iacs,uscg_classification,ferns,chopra,deutsche_flagge,emisa,iadc,solas,isps,maritime_cyber_risk,imo_guidelines, cyber_guidelines_v5} to ground this overview, and enrich it with insights from participant interviews.} Additionally, we discuss prior research on operator studies and maritime cybersecurity, situating 
this study within the broader research landscape.

\subsection{Marine Context}

\noindent\textbf{Ships and Mariners.} Maritime vessels serve diverse roles within global infrastructure, encompassing cargo ships, passenger ships, and specialized vessels such as research ships, tugboats, and buoy tenders. For the purposes of this study, \emph{ships} are defined as professionally operated vessels engaged in commercial, scientific, or military operations, excluding recreational boats, yachts, and other privately owned or non-industrial watercraft. These industrial ships often blend civilian and military contexts; for example, the Military Sealift Command (MSC) employs civilian mariners to transport military cargo. The varied ship types and their operations highlight the complexity of maritime activities and the cybersecurity challenges they face. This study reflects this diversity by including participants with experience across different ship types, showcasing the blended civilian and military nature of the maritime industry.

\new{We define \emph{mariners} as credentialed professionals responsible for the safe and effective operation of ships.} Key roles include captains, chief mates, and engineers, whose responsibilities extend to ensuring cybersecurity. This study focuses on officer-level mariners who oversee critical systems, offering valuable insights into cybersecurity practices in maritime operations. Further details on ship classifications and mariner roles are provided in~\Cref{sec:ship_details}.

\noindent\textbf{Regulatory Frameworks.} Maritime operations are governed by a multilayered regulatory framework. The International Maritime Organization (IMO) establishes global conventions, while classification societies (e.g., Lloyd’s Register, ABS) set technical standards. National agencies like the U.S. Coast Guard enforce compliance, and flag states oversee vessels registered under their jurisdiction. This fragmented landscape creates overlapping and sometimes conflicting requirements, complicating efforts to address cybersecurity \new{risks} comprehensively. \new{Hence, there is a pressing} need for unified regulations tailored to modern maritime threats.

\new{\noindent\textbf{IT/OT Responsibilities.} Cybersecurity in maritime operations involves protecting both Information Technology (IT) and Operational Technology (OT) systems from threats that can compromise vessel safety, cargo, crew, and environmental integrity. IT systems manage data exchange and business functions, while OT systems control physical processes such as engine operation, navigation, and cargo handling. Unlike other critical infrastructure sectors (e.g., energy), where IT and OT responsibilities are typically distributed across specialized teams, mariners are often responsible for both domains~\cite{nist80082r3,imo_guidelines}. This includes managing Internet access and administrative software (IT), as well as monitoring propulsion, navigation, and cargo systems (OT), often without dedicated cybersecurity support or real-time assistance from shore-based teams.

\noindent\textbf{Cybersecurity Risks.} Based on IMO and BIMCO (Baltic and International Maritime Council)-aligned frameworks~\cite{imo_guidelines,cyber_guidelines_v5}, maritime cyber risks can be grouped into the following key categories:\vspace{-\baselineskip}
\begin{itemize}[leftmargin=*,noitemsep]
  \item \textbf{Bridge and navigation systems:} Systems such as ECDIS (Electronic Chart Display and Information System), AIS (Automatic Identification System), GPS (Global Positioning System), and radar are vulnerable to tampering, GPS spoofing, malware infections, or outdated software, potentially impacting vessel routing and situational awareness.
  
  \item \textbf{Propulsion and power control:} Digital systems managing engines and power distribution are often integrated with shore-based monitoring tools, exposing them to remote access risks or accidental disruption via unsafe updates.

  \item \textbf{Cargo handling systems:} These interface with port terminals and rely on data integrity for stowage plans and \cam{manifests}, making them targets for manipulation or denial-of-service attacks.

  \item \textbf{Communication and administrative systems:} Email servers, VSAT links, and crew internet access points are entry vectors for phishing, malware, and ransomware, especially when software patching and access controls are inadequate.

  \item \textbf{Access control and surveillance:} Digital security systems including CCTV, gangway monitors, and electronic “personnel-on-board” logs may be compromised to mask unauthorized access or disrupt safety protocols.

  \item \textbf{Third-party and supply chain exposure:} Risks may arise from vendors, contractors, and even port officials who connect external devices or use removable media onboard, often without strict \cam{vetting}. These entry points can be used to pivot laterally between IT and OT environments if segmentation is weak.

  \item \textbf{Human error and social engineering:} Crew members may unintentionally introduce risks through phishing, poor password practices, or unsanctioned software use. Social engineering and spear-phishing remain persistent threats.
\end{itemize}

These threats are exacerbated by maritime-specific challenges, including long software maintenance cycles, limited onboard IT support, legacy equipment, and fragmented regulatory oversight\cam{~\cite{cyber_guidelines_v5}}. Unlike land-based critical infrastructure, mariners often operate these systems in isolation while at sea, without real-time cybersecurity assistance. While these categories provide a useful starting point, they primarily reflect top-down, risk management perspectives. Our study complements this view with a bottom-up lens, drawing from mariners’ firsthand accounts. In Section~\ref{sec:categorization}, we present an empirically grounded categorization of threats based on participant experiences, which expands on these official categories by capturing how mariners perceive threat actors, vulnerable moments, and operational impacts.}

\subsection{Related Work}

\noindent\textbf{Operator Studies.} Prior research has explored the cybersecurity practices and \new{perceptions} of various operators, such as web administrators, malware analysts, and bug hunters. These studies use qualitative methods like semi-structured interviews to gain detailed insights into workflows, challenges, and decision-making models, which align with our chosen methodology for this study~\cite{sahin2023investigating,yong2021inside,wong2024comparing,akgul2023bug,votipka2018hackers}. \new{More recent work has also investigated cross-domain collaboration between cyber and OT experts in energy infrastructure, revealing cultural and epistemic differences between domains and calling for interdisciplinary approaches to impact assessment~\cite{gallardo2024interdisciplinary}.  Similarly, user studies in industrial control systems (ICS) contexts report significant cybersecurity challenges, such as organizational barriers to OT security culture~\cite{evripidou2023exploring}, usability problems in PLC security features~\cite{li2024windows98}, and calls for user-centered design of industrial security solutions~\cite{nunes2024exploiting}, with asset owners echoing many of these issues~\cite{fla2024challenges} and mindset differences further shaping OT practitioners' cybersecurity perceptions~\cite{evripidou2022understanding}. While these works inform the broader understanding of operator perspectives, they largely focus on digital or stationary critical infrastructure contexts. In contrast, our study examines mariners as mobile CPS operators responsible for both IT and OT systems in high-risk, high-autonomy environments. This presents unique operational and environmental constraints that shape mariner perceptions and behaviors related to cybersecurity.}%

\noindent\textbf{Maritime Cybersecurity.} Research on maritime cybersecurity is largely underexplored, with most studies focusing on isolated aspects of the field. For example, the vulnerabilities of Very Small Aperture Terminals (VSATs), used for shipboard communication, have been highlighted~\cite{pavur2020tale}, along with defenses against GPS spoofing attacks~\cite{liu2021stars}. Theoretical explorations of potential maritime attacks and preliminary analyses of navigation systems further underscore the need for comprehensive research~\cite{progoulakis2021cyber,direnzo2015little,tran2021marine,amro2022navigation}. These works primarily address technical vulnerabilities without examining how mariners engage with systems or perceive cybersecurity risks. By providing empirical insights into mariners’ cybersecurity practices and \new{perceptions}, this study connects technical findings to the human factors critical for securing maritime operations.

\section{Method}

\label{sec:method}

We interviewed 21 participants for our main study and 2 participants during a pilot study to explore the maritime cybersecurity landscape and address our research questions. \new{The study followed} best practices, including obtaining approval from our organization’s Institutional Review Board (IRB), conducting a pilot study to refine our methods, ensuring saturation to determine the study’s completion, and calculating inter-coder agreement to ensure reliability.

\subsection{Interview Design}

The following section outlines \new{our} final interview design, refined through pilot study insights, and justifies the \new{selected} questions.

\subsubsection{Interview Questions} Each set of questions was designed to address specific aspects of our research questions (RQs), ensuring a comprehensive exploration of mariners’ \new{perceptions of cybersecurity} (RQ1) and practices (RQ2). \cam{We also drew on Cyber-Informed Engineering (CIE) principles~\cite{lampe2024application} to inform the operational framing of our questions.}\\
\textbf{General Security Questions.} These questions aimed to establish participants’ general perceptions of security and served as a foundation for discussing more specific topics. They were designed to align with RQ1 by encouraging participants to reflect on threats to the ship and its operations, identify the types of threats that most concerned them, and consider how security threats and practices varied depending on location, personnel, or operational context. \\
\textbf{Cybersecurity Practices and Incidents Questions.} These questions directly addressed aspects of both RQ1 and RQ2. For RQ1, they explored participants’ \new{perceptions of cybersecurity}, including their understanding of cybersecurity concepts and concerns about cyber threats. For RQ2, these questions examined mariners’ cybersecurity practices and behaviors, including their confidence in managing incidents and the adequacy of their training. This included establishing participants’ baseline understanding of cybersecurity, exploring their experiences with cyber-attacks, uncovering their primary concerns regarding cybersecurity threats, and examining their training and preparedness. \\
\textbf{Comparative Cybersecurity Questions.} These questions provided additional insights into RQ1 by bridging participants’ understanding of cybersecurity and traditional security (e.g., physical threats such as piracy or unauthorized access). This comparative approach assessed whether and how mariners’ \new{perceptions of cybersecurity} differed from their views on traditional security threats. These questions investigated participants’ perceptions of cybersecurity threats, identified the threats that most concerned them, and compared these perspectives with their views on traditional security threats. \\
\textbf{Regulation and Standards Questions.} These questions were central to RQ2, focusing on how standards and regulations influence mariners’ cybersecurity practices. They examined participants’ views on existing safety and security standards and their perceptions of emerging cybersecurity regulations. The focus was on understanding participants' assessment of current standards, whether their perceptions of general safety and security standards differed from those of cybersecurity standards, and how they viewed the adequacy and implementation of these regulations. \\
The full set of interview questions can be found in~\Cref{sec:interview}.

\subsubsection{Pilot Study}
Our pilot study, conducted in two rounds, refined the interview question set to better capture mariners’ \new{perceptions of cybersecurity} and practices. The first round added broader security questions to address participants’ tendency to focus narrowly on technical threats like malware and phishing, encouraging them to consider non-digital threats such as physical access vulnerabilities. The second round built on this by introducing mirrored cybersecurity questions to align with the broader security topics, ensuring balanced and comprehensive coverage of both perspectives. While insights from the second pilot informed the final question set and were included in the analysis, saturation calculations began with Participant 1 in the main study. We describe this process in much more detail in~\Cref{sec:pilot}.

\subsection{Recruitment, Survey, and Interviews}
Participants were recruited through multiple channels, including LinkedIn \new{(1 participant)}, Reddit \new{(1 participant)}, gCaptain \new{(8 participants)}, and personal connections\new{/snowballing (11 participants)}. Among \new{the online recruitment methods}, gCaptain proved to be the most effective in attracting participants. Recruitment in this field presented unique challenges due to the demanding work schedules of mariners, a general mistrust of unsolicited online contacts, and limited internet access for many potential participants. These barriers necessitated sustained and deliberate efforts to ensure sufficient participants. Despite these challenges, we successfully recruited participants by emphasizing the study’s relevance and importance. \new{Participants were not compensated for their participation in this study; see ~\Cref{sec:ethics} for rationale and further discussion.}

To assess participant suitability for interviews, we utilized a Microsoft Forms survey to collect background information (see~\Cref{sec:survey}). \new{Eligibility for participation required holding an officer position in shipping.} Of the 32 individuals who completed the survey, 30 \new{met this criterion} and were contacted for an interview. Ultimately, 23 of these individuals scheduled interviews, with 2 participating in the pilot study and 21 contributing to the main study. To expand the participant pool, we incorporated snowball sampling after the first five interviews. The survey remained open for three months.

The interviews were conducted via Zoom, an online video conferencing platform, and recorded with participants’ consent. They lasted 60 minutes on average, though some extended beyond this due to participants having a lot to share, underscoring the depth of engagement. This flexible approach ensured participants could fully express their thoughts without time constraints.

\subsection{Participant Demographics}

\begin{table}[ht]
\vspace{1em} %
\centering
\small
\begin{tabular}{|p{0.45cm}|c|c|p{1.51cm}|p{1.57cm}|p{0.9cm}|}
\hline
\textbf{$^\diamond$P, $^\triangle$PP} & \textbf{$^\bullet$YiS} & \textbf{Position} & \textbf{Ship Type(s)} & \textbf{Goods Transported} & \textbf{Crew} \textbf{Size} \\ \hline
PP1 & 9 & 2nd Eng. & Tanker & Containers, LNG & 25 \\ \hline
PP2 & 22.5 & 2nd Mate & Tanker & Refined Petro Products & 8 \\ \hline
P1 & 8 & 3rd Eng. & Tanker & Asphalt, HFO & 15 \\ \hline
P2 & 5 & Chief Mate & Tanker & Refined Petro Products & 20 \\ \hline
P3 & 2 & 3rd Eng. & Tanker & Containers, LNG & 22-35 \\ \hline
P4 & 25 & Captain & Dry Cargo & Personnel, Equipment & 21 \\ \hline
P5 & 15 & Chief Eng. & Dry Cargo & Containers & 21 \\ \hline
P6 & 28 & Captain & Cargo, Tanker,\ \ \ \ \ \ \ \ Research & Containers, LNG, Bulk & 7-240 \\ \hline
P7 & 30 & Captain & Dry Cargo & Containers & 20 \\ \hline
P8 & 2 & $^\circ$DWO & Military & - & 50 \\ \hline
P9 & 8 & 1st Mate & Sub-sea \ \ \ Construction & Fiber-optic Cable & 55 \\ \hline
P10 & 35 & Chief Mate & Military & Military & 28 \\ \hline
P11 & 37 & Captain & $^\dagger$MSC & Military & 150-200 \\ \hline
P12 & 2 & $^\circ$DWO & Military & - & 50 \\ \hline
P13 & 3 & $^\ddagger$OO & Military & - & 24 \\ \hline
P14 & 26 & Captain & Research & Personnel, Equipment & 16 \\ \hline
P15 & 10 & $^\times$EO & Passenger & People & 1,800-2,200 \\ \hline
P16 & 39 & Captain & $^\dagger$MSC & Military & 120 \\ \hline
P17 & 5 & Chief Mate & Sub-sea \ \ \ Construction & Fiber-optic Cable & 60 \\ \hline
P18 & 1 & 3rd Mate & Military & Military & 83 \\ \hline
P19 & 18 & Captain & Dry Cargo & Containers & 24 \\ \hline
P20 & 14 & $^*$PHO & Passenger & People & 1,500 \\ \hline
P21 & 15 & Chief Mate & Dry Cargo & Containers, Military & 19-35 \\ \hline
\end{tabular}
\vspace{1em} %
\caption{Participant Information}

{$^\diamond$Participant;
$^\triangle$Pilot Participant; 
$^\bullet$Years in Shipping
$^\circ$Deck Watch Officer; 
$^\dagger$Military Sealift Command; 
$^*$Public Health Officer; 
$^\ddagger$Operations Officer; 
$^\times$Environmental Officer;}

\label{tab:participant_info}
\end{table}

Participant demographics are summarized in Table ~\ref{tab:participant_info}, providing an overview of the individuals included in this study. Mariners interviewed had varying levels of experience, ranging from 1 year to 39 years on the job. The average crew size varied significantly depending on the type of vessel, with research vessels hosting the smallest crews and passenger vessels the largest. \new{Additionally, participants represented a wide age range, with 2 aged 20–29, 5 aged 30–39, 3 aged 40–49, 2 aged 50-59, 1 aged 60–69, and 1 aged 70–79; 9 participants did not share their age.}

A diverse range of affiliations were also represented in the participant pool. Nine participants indicated some level of military affiliation, including reservist mariners working on civilian ships, mariners working exclusively in military roles, and civilians employed by the Military Sealift Command (MSC) transporting military cargo. Participants reported sailing under a wide range of flag states, spanning North America, Europe, Asia, Africa, and Oceania. These included major flag states such as the United States, Panama, Malta, and Liberia, as well as several others, reflecting the global nature of the maritime industry. \new{Flag states indicate the regulatory environments under which participants were trained and operated, serving as a kind of regulatory “home country.” Additional details on participant flag states are provided in~\Cref{sec:demo_data}.}

The study included 18 male participants and 3 female participants. This is notable given that, according to the International Maritime Organization (IMO), women made up only 1.2\% of the global seafarer workforce in 2021 \cite{women_in_maritime}. With 14\% of our participants identifying as women, this study incorporates a higher proportion of women than the industry average. This greater representation enriches the study by capturing a broader range of experiences and perspectives, which may otherwise be underrepresented in maritime cybersecurity research.  However, due to participant concerns about anonymity and the low percentage of women in the maritime industry, we refrain from identifying participants by gender in Table ~\ref{tab:participant_info}. The diversity in the mariners’ experience, vessel affiliations, and roles, combined with achieving saturation, underscores the robustness of this study and the quality of its findings.

\subsection{Data Collection and Analysis}
All interviews were recorded, transcribed, and anonymized. Transcriptions were securely stored and accessible only to researchers approved by the Institutional Review Board (IRB). The primary interviewer manually edited each transcript, cross-referencing the audio recordings to validate the accuracy of the information and highlight key interview questions.

\new{We employed an iterative open coding methodology~\cite{srivastava2009practical}, refining the codebook as data was analyzed. Two coders independently coded a transcript, then met to reconcile differences and update the codebook. This process repeated until the final codebook was established, which is hosted at an \cam{Open Science Framework (OSF) repository} along with associated definitions~\cite{osf_codebook}}. From the 30 interview questions and subquestions, our analysis produced 448 codes in total. \new{Intercoder reliability was assessed using Cohen's Kappa \cite{fleiss2013statistical}, yielding a score of 0.837. This metric demonstrated excellent reliability and confirmed consistent extraction of themes across coders.}

\new{Saturation was used to determine when a sufficient number of interviews had been conducted. Following Guest et al.~\cite{guest2020simple}, we considered saturation reached when new themes ceased to emerge and the proportion of new codes fell below 10\%, ultimately reaching 0\% in the final set of interviews.} Given the extensive diversity within the maritime domain—including variations in ship type, affiliations (military and civilian), and onboard roles—we conducted an additional five interviews beyond saturation to ensure comprehensive representation and confidence in the findings.

\subsection{Ethics}
\label{sec:ethics}
This study followed standard ethical practices to ensure the rights, privacy, and safety of participants were respected throughout the research process. Institutional Review Board (IRB) approval was obtained prior to initiating the study. Participant data was anonymized and securely stored, with access restricted to IRB-approved researchers involved in the project. \cam{Participant emails were stored separately to maintain contact with participants and share study findings; no identifiers were linked to research data.} Consent procedures were carefully implemented to ensure informed participation. Participants provided explicit consent before completing the initial survey. Before each interview, consent was reaffirmed verbally, and participants were reminded of their right to skip any questions they did not wish to answer.

\new{Participants did not receive compensation, a decision approved by our IRB and aligned with prior work involving professional operators, which similarly did not offer compensation~\cite{sahin2023investigating,6046004,219400,7958576,205176,derr2017keep,voronkov2019system,haney2018s,gorski2018developers,gerlitz2021please}. Participation in our study did not pose financial burdens. In such cases, compensation is not ethically mandatory, and in fact, entails increased methodological risks (e.g., participation coercion, biased participant motivations)~\cite{bruckman2024compensation}. These risks are especially pronounced when working with high-income populations, where compensation would need to be proportionally higher to be meaningful, potentially increasing coercive pressure. We also note that most participants were employed by organizations that prohibit accepting external payments.}

By adhering to established ethical guidelines, including \cam{informed consent} and secure data handling, this study ensured participant confidentiality and upheld the principles of responsible research.
\subsection{Limitations}
As with any qualitative user study, this research has inherent limitations that should be considered when interpreting the findings. One limitation is the potential for social desirability bias, a common challenge in qualitative research. This occurs when participants provide responses they believe are expected or socially acceptable rather than their genuine opinions. To mitigate this, we carefully designed neutral interview questions and emphasized to participants that there were no “right” answers. The interviewer also encouraged honest and candid responses. 

\new{It is possible that for some participants, some of our prompts may have been unintentionally leading or inaccurately assumed baseline familiarity with cybersecurity concepts (e.g., Q10, Q16, and Q19).
We attempted to mitigate this concern through piloting our questions and designing them to be open-ended and flexible, while also allowing for clarifications through our semi-structured format. However, it remains a core limitation of interview-based research where question framing can influence responses.}

While external validity is inherently limited in qualitative studies, we took steps to enhance the robustness of our findings. This included collecting data from a diverse range of participants and ensuring saturation. These measures provide a strong foundation for understanding mariners’ perspectives and highlight critical insights into cybersecurity practices and challenges. The goal of such qualitative research is not to be fully comprehensive or generalizable but to uncover rich, nuanced insights in a relatively unexplored domain, offering a foundation for future studies in this space.

\section{Cybersecurity Perceptions}
This section presents findings related to RQ1: \emph{What are mariners’ \new{perceptions of cybersecurity}?} Insights into mariners’ \new{perceptions} were derived from responses to the general security, comparative cybersecurity, and cybersecurity practices and incidents questions described in our methodology (\Cref{sec:method}). While most results tie directly to these questions, certain emergent themes, such as concerns about autonomous ships, arose unprompted during discussions on cybersecurity. Each bolded subsection below (i.e., \new{4}.x) outlines the specific questions or themes that guided the findings. In the second subsection, we present a \new{categorization} of cyber threats derived from mariner interviews. In conjunction, these insights and the proposed \new{categorization} give a holistic view of how participants view maritime cybersecurity and the factors that shape their perceptions. 

\subsection{Investigating Mariner \new{Cybersecurity Perceptions}}
\label{sec:mental_model}
This subsection highlights insights into mariner \new{perceptions of cybersecurity}. What do mariners perceive as security threats, and to what extent do they include cybersecurity in this perception? How do these views diverge from more conventional understandings of security? Furthermore, we examine whether mariners’ experiences with cyber incidents have influenced these \new{perceptions}.

The following bolded insights explore key themes derived from participant responses to the general security, comparative cybersecurity, and cybersecurity incident interview questions. We first address foundational perceptions and misconceptions shaping mariners’ \new{perceptions} before discussing specific threats and emerging concerns raised during the interviews.

\noindent\textbf{\blackcircled{A} \new{Difference in Perceived Physical and Cyber Impacts.}}
This insight emerged from comparing responses to the general security and comparative cybersecurity interview questions. Participants were asked to reflect on the threats they associate with security and whether those included cybersecurity. \new{In many cases, mariners described physical threats (e.g., piracy, terrorism) more readily than cyber threats, possibly because they are less visible or tangible.}

When discussing physical security, only three participants explicitly included cyber threats in their responses\new{, although all} 21 mariners mentioned risks such as piracy, terrorism, and physical harm. \new{Participants did not appear to associate threats like ransomware or system sabotage with the same level of urgency as physical attacks.} \cam{This likely reflects both limited familiarity with digital threats and a practical focus on the immediate dangers of maritime work. As a result,} mariners may place less emphasis on cybersecurity practices, such as verifying email authenticity or updating navigation software, especially under demanding workloads.

\new{These realities can be} compounded by \new{fatigue and extended work shifts, which reduce overall} vigilance. One participant noted, \emph{“A lot of errors that are made, it’s due to fatigue”} (P5). Another elaborated, \emph{“If you’re working 12 hours a day for 90 days, you don’t have anything left… After 30, 40 days, you’re not as alert and you just don’t care”} (P6). Under such conditions, \new{cyber incidents may go unnoticed for longer periods}, aligning with concerns expressed by another mariner: \emph{“If somebody was smart enough… They could bring the maritime world to a crawl… it’s probably a matter of time”} (P9).

While some participants \new{did link} physical and cyber threats\new{--citing spoofing of navigation equipment, for example--}others focused on the immediate, tangible dangers of maritime operations. \new{O}ne mariner recounted experiencing \emph{“three attempted piracies”} while onboard (P7), \new{illustrating why physical threats remain top-of-mind. Better} integrating cybersecurity awareness \new{and practical countermeasures} into existing security frameworks \new{could help mariners see how digital attacks, too, can lead to monetary loss, operational disruption, or risks to personnel safety.} \\
\noindent\textbf{\blackcircled{B} Military Perspectives on Cybersecurity.} While the previous insight highlights a broad \new{difference}, some mariners demonstrated unique perspectives informed by military experience, including the Military Sealift Command (MSC), a civilian branch of the U.S. Navy. This theme also arose from the general security and comparative cybersecurity interview questions.

Seven out of the nine participants with MSC or Coast Guard experience exhibited a nuanced understanding of cyber-attacks, linking them not only to computers but also to navigation equipment and other critical systems. This perspective contrasts with other participants who primarily associated cyber-attacks with Information Technology (IT) systems, such as administrative networks or email servers. However, even with this expanded awareness, these mariners \new{tended to emphasize day-to-day operational concerns over the possible physical impacts of cyber threats}.

One MSC-affiliated mariner explained their confidence in traditional navigation methods: \emph{“I’m very, very comfortable navigating without anything other than a sextant, a stopwatch, a chronometer, and a paper chart. Most MSC Officers are trained that way”} (P11). This reflects how military training shapes mariners’ \new{perceptions}, emphasizing operational resilience and alternative navigation. \\
\noindent\textbf{\blackcircled{C} Cyber-Attack Experience and Impact.} 
Participant responses further illustrate how direct experiences with cyber-attacks influence their perspective. These findings stemmed from the cybersecurity practices and incidents interview questions. 

Participants recounted their cyber-attack experiences, particularly GPS, \cam{AIS}, and radar spoofing. In total, 10 mariners described direct encounters with such incidents. For instance, one participant reported being \emph{“cyber-attacked by Iran”} with their ship moved into Iranian waters, adding, \emph{“I’ve been viciously spoofed”} (P2). Another mariner recalled experiencing AIS spoofing near Taiwan: \emph{“It was very unnerving… we operate so long on AIS that it becomes… a source of truth… we had to convince ourselves that it wasn’t real, and it took a concerted effort to do that”} (P4). Despite these encounters, these participants \new{did not always describe connections between cyber-attacks and} potential physical consequences, \new{ suggesting these risks may not be fully integrated into their broader view of security.}

\noindent\textbf{\blackcircled{D} High-Impact Cyber Threats.}
Building on mariners’ experiences, we next examine key threats they identified, particularly from phishing, physical access, and remotely monitored equipment. These insights emerged from the cybersecurity practices and incidents interview questions, where participants discussed what they viewed as the most significant cybersecurity concerns and why.

Mariners are particularly vulnerable to phishing threats due to their reliance on email for nearly all business communication, including with companies and ports. Overall, 10 participants expressed this concern for email and phishing threats. One participant observed, \emph{“You get an email every day from the office… hundreds of emails a day”} (P10). The high volume, coupled with long hours at sea and infrequent access to personal devices, can lead to mistakes. Another mariner shared, \emph{“It always seems to come in by email… You clicked the wrong button and game on”} (P6).

Ships are also at risk from physical access threats, as many third-party contractors and technicians board the ship during port stays. 11 participants mentioned concerns over threats from third parties like this. Mariners highlighted the difficulty in verifying these individuals as well. One said, \emph{“During the port time… that’s when strangers could easily access the ship”} (P1), while another admitted, \emph{“I’ve had contractors ask me to stick their USBs into printers… we have to trust them”} (P2).

Furthermore, 13 participants mentioned a growing concern involving remotely monitored equipment, such as engines, which are increasingly connected to shore-based systems. They expressed that this connectivity creates vulnerabilities that they are not equipped to address. One participant explained, \emph{“I think engine monitoring equipment is our bigger threat… you’re putting that capability out there into the world for it to be hacked”} (P15). Another shared, \emph{“The engines are controlled by a computer hooked up to the Internet… someone could conceivably just completely run amok”} (P16). \\
\noindent\textbf{\blackcircled{E} Cyber Threat Misconceptions.} Despite identifying key threats, many mariners still hold misconceptions about system vulnerabilities. These misconceptions, uncovered through the cybersecurity practices and incidents questions, include \new{limited familiarity with} data interchange and system connectivity risks.

Six mariners \new{indicated limited recognition of} how interconnected systems can introduce vulnerabilities. For example, one participant believed, \emph{“Our systems are pretty secure because we only use USBs issued by the company”} (P9). This reflects a common misunderstanding, as reliance on trusted devices does not account for the broader attack vectors introduced by system interconnectivity and third-party vulnerabilities. However, other participants offered a more nuanced view of these risks. As one mariner aptly observed, \emph{“That constant interchange of data… every single time there’s an interchange of data, there’s a potential for a threat to come on board”} (P7). \new{This contrast underscores differences in mariners’ awareness and the need for training to bridge that divide.}

A sense of confidence in the open ocean also emerged as a common theme. While 13 mariners reported feeling safer at sea, viewing it as a reprieve from physical security threats like piracy, this sense of security often overlooks cyber risks. One participant described the mindset: \emph{“That’s the time when we are actually relaxed… we can focus on our work”} (P1). However, another highlighted the vulnerability of being far from assistance, stating, \emph{“At sea, maybe the attack surface is less compared to shore. However, no one can help you”} (P3). \new{Such} confidence could leave ships \new{underprepared} for cyber-attacks, particularly with emerging technologies like Starlink increasing connectivity in remote areas.

\noindent\textbf{\blackcircled{F} Emerging Concerns over Autonomous Ships.}
In addition to the themes that explicitly surfaced from interview questions, participants frequently raised concerns about future technologies, particularly autonomous ships. Six participants expressed concerns about increased cybersecurity risks with the adoption of autonomous and remotely operated vessels. As these vessels rely heavily on interconnected systems, the potential for cyber threats increases drastically. One participant remarked, \emph{“If you wanted an autonomous ship, you would have to worry about everything because everything has to be connected”} (P2). This connectivity, while essential for automation, creates multiple entry points for potential cyberattacks.

The shift toward remote operations underscores the urgency of these concerns. A participant shared their experience: \emph{“The operation I’m at now, we’re actually driving our vessel remotely… if I can operate the ship’s heading, the nav system, and the power plant from a thousand miles away, so can somebody else”} (P4). This insight highlights how the very systems enabling remote control could also be exploited by malicious actors. As automation becomes a reality in the maritime industry, ensuring robust cybersecurity measures is critical to safeguarding these advanced systems from potentially catastrophic threats.

\begin{tcolorbox}[insightbox]
 \textbf{Mariners often prioritize physical threats over cyber risks, reflecting gaps in their \new{perceptions of cybersecurity}. While some participants’ experiences with cyberattacks shaped their awareness, misconceptions about system vulnerabilities and misplaced confidence in open ocean safety persist.}
\end{tcolorbox}

\subsection{\new{Categorization} of Cyber Threats}
\label{sec:categorization}

\begin{figure*}[htb!]
    \centering
    \includegraphics[width=\textwidth]{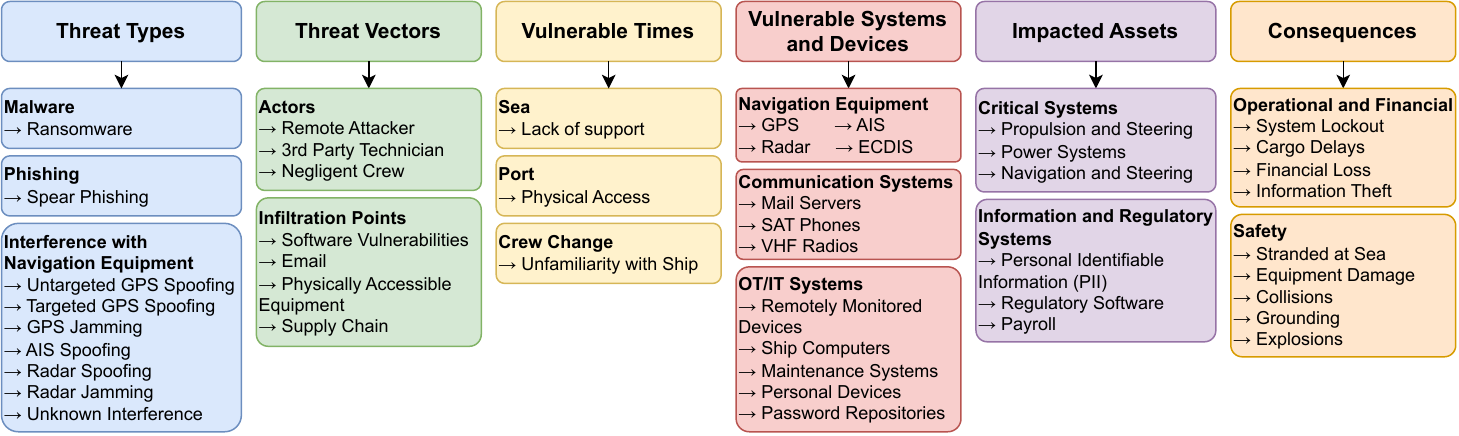} %
    \caption{\new{Categorization} of Cyber Threats}
    \label{fig:categorization_paper}
\end{figure*}

\new{This categorization of results, shown in Figure~\ref{fig:categorization_paper}, summarizes the threats described in participants’ responses to the cybersecurity comparison, practice, and incident questions. It reflects mariners’ perceptions of cyber risks grounded in lived experience. As discussed in Section~\ref{sec:background}, prior frameworks from IMO and BIMCO organize risks by shipboard system (e.g., bridge, propulsion)~\cite{imo_guidelines,cyber_guidelines_v5}. In contrast, our categorization groups threats by type, entry point, timing, and impact—dimensions that reflect how mariners actually experience cyber risk. This structure surfaces scenarios like crew changeover or remote equipment monitoring, which cut across multiple technical systems but are rarely emphasized in official frameworks. This bottom-up view is not intended to be comprehensive but offers a practical complement to top-down frameworks by centering the operational realities and concerns of frontline personnel. It can help guide future research, regulatory development, and training to better align with mariner needs. Additionally, please note that some participants reported experiences across multiple categories, so totals may exceed the number of interviewees.}

\subsubsection*{Threat Types}
This category outlines cyber threats participants either experienced or feared could impact their ship.

\noindent\textbf{Malware.} Malware and particularly ransomware were mentioned as cyber threats by eight participants. They were concerned these types of cyber threats could take down ship and port systems, like payroll and regulatory programs. Four participants mentioned experiencing this class of cyber-attack on ship.

\noindent\textbf{Phishing.} Phishing and particularly spear phishing (more targeted phishing attacks) were mentioned as cyber threats by 10 participants. The largest shipping company in the world, Maersk, experienced a cyber-attack via spear phishing \cite{greig2024ransomware}, and some participants were on ship at the time of the attack. For those participants, such incidents highlighted the tangible risks posed by cyber threats. Two participants mentioned experiencing this class of attack on ship.

\noindent\textbf{Interference with Navigation Equipment.} Furthermore, 10 participants experienced all ranges of cyber-attacks due to interference with navigation equipment. This included untargeted GPS spoofing (2 participants), targeted GPS spoofing (2 participants), GPS jamming (3 participants), AIS spoofing (3 participants), radar spoofing (1 participant), and radar jamming (1 participant). Two participants also mentioned "unknown interference" with navigation equipment where they thought a cyber-attack occurred but could not verify it.

\subsubsection*{Threat Vectors}
This category details actors and mechanisms that could be exploited to infiltrate systems and cause threats.

\noindent\textbf{Actors.} Overall, 18 participants expressed concern about actors contributing directly or indirectly to cyber threats. These included remote attackers targeting off-ship equipment (13 participants), negligent crew lacking or disregarding cyber hygiene practices (4 participants), and third-party technicians with unfettered ship access requiring inherent but unverifiable trust (5 participants).

\noindent\textbf{Infiltration Points.} 16 participants mentioned concern over physical and digital pathways that enable attackers to exploit systems. Three participants worried about software vulnerabilities leading to cyber threats on ship. With the highly publicized phishing attacks on shipping systems and cybersecurity training focus, they also worried about email as an infiltration point (10 participants). Additionally, 11 participants mentioned physically accessible equipment as a vector for introducing cyber threats. This could be due to open USB ports on equipment like their \cam{ECDIS} or being unable to secure on-ship computers from third parties that enter the ship. Finally, five participants mentioned supply chain vulnerabilities leading to ship equipment with malicious alterations integrated during manufacturing (e.g., navigation equipment or ship computers). They worried that with the diversity of equipment types, vendors, and manufacturers, someone could target their equipment before \new{installation}. This worry was compounded by the fact that many ships regularly purchase new electronic parts to replace old ones and send orders via email, which they also believed to be insecure.

\subsubsection*{Vulnerabilities}
This category covers scenarios, systems, and devices identified as vulnerable to threats.

\noindent\textbf{Vulnerable Times.} From 11 participants, we heard of three distinct vulnerable scenarios when on ship. The first was at sea, very far from civilization, because they could not get immediate support from shoreside staff and law enforcement (5 participants). The second scenario was at port because so many third parties had physical access to equipment (7 participants). The final scenario was during crew change because the new crew could be unfamiliar with the ship, leading to poor cybersecurity practices and the exploitation of vulnerabilities by attackers (1 participant). Although these scenarios span most operational periods, each has distinct sources of vulnerability that require specific mitigation strategies.

\noindent\textbf{Vulnerable Systems and Devices.} In total, 16 participants mentioned concerns over vulnerabilities in three types of systems and devices. The first was navigation equipment, including GPS, AIS, Radar, and ECDIS (8 participants). The second was communication systems, including mail servers, satellite phones, and VHF radios (3 participants). The third were OT and IT systems, including remotely monitored devices (e.g., engines), ship computers, maintenance systems, personal devices, and password repositories (11 participants).

\subsubsection*{Consequences and Impacts}
This category describes the impacts on assets and the broader consequences of cyber threats.

\noindent\textbf{Impacted Assets.} We heard concerns from 11 participants over two types of assets that could be impacted by cyber threats. The first asset was critical equipment, including propulsion and steering systems, power systems, and navigation equipment (9 participants). The second asset was information and regulatory equipment, including regulatory software, personal identifiable information (PII) storage, and payroll systems (3 participants).

\noindent\textbf{Consequences.} We had 12 participants mention two categories of consequences to the aforementioned cyber threats. The first category was operational and financial, including system lockout, cargo delays, financial loss, and information theft (3 participants). The second category was safety, including becoming stranded at sea, damage to equipment, collisions with other ships or structures, grounding, and even explosions due to carrying hazardous materials on ship (e.g., liquified natural gas), which included 10 participants.

\section{Shaping Practice: Training and Regulation}

This section presents the findings addressing RQ2: \emph{What are mariners’ cybersecurity practices, and what role do training and regulations play in shaping these practices?} Insights are drawn from participants’ responses to questions specifically addressing their experiences with cybersecurity, cyber incidents, and regulations. The analysis aims to highlight the practical challenges mariners face and the systemic gaps that affect their cybersecurity practices.

The findings are structured into two subsections. The first explores mariners’ training, behaviors, and practices, giving insight into how training content translates—or fails to translate—into effective preventative and reactive cybersecurity measures. The second examines mariners’ experiences with maritime cybersecurity regulations, identifying areas of alignment and disconnect between regulatory frameworks and practical shipboard operations.

\subsection{Mariner Cybersecurity Training, Practices, and Behaviors} 
\label{sec:training_practices}
This subsection addresses mariners’ cybersecurity practices, focusing on how training influences their preventative and reactive behaviors. Insights were derived from participants’ responses to questions about their cybersecurity training, their experiences with cyber incidents, and their daily cybersecurity practices. Additionally, emergent themes, such as the management of new technologies like Starlink, arose organically during discussions.

\noindent\textbf{\blackcircled{A} \new{Basic} Cybersecurity Training.}
Participants were asked about the cybersecurity training they received to assess its scope and effectiveness. All 21 mariners described their cybersecurity training as primarily \new{focused} on email and USB threats. This perspective was consistent across participants from both industry and military sectors. The training was perceived as generic, with little adaptation to the unique challenges faced by mariners. As one participant noted, \emph{“I’ve been in this industry for a long time, and I know a lot of folks. That cyber security section for most companies… is just boilerplate… we’ve met the requirement by speaking to it”} (P4). This approach \new{may not fully prepare mariners for the range of possible cybersecurity scenarios}, leaving mariners under-equipped to address modern maritime cyber threats. \\
\noindent\textbf{\blackcircled{B} Cybersecurity Training: \new{Limited} IT and OT \new{Integration}.}
While participants were not directly asked about the relationship between IT and OT security, the \new{gap} between these domains emerged when participants described their training content. On ships, these systems are increasingly interconnected, with IT networks often serving as entry points for attacks targeting OT systems.

Some mariners explicitly noted the omission of OT-related risks in training. One participant highlighted this gap: \emph{“They typically focus on … IT security. However, if we talk about … maritime, we should also consider … OT security. The seafarers, or the office staff, are typically not even aware of the cyber risk of OT components, and in such an attack of course they cannot know what [to] do”} (P3). \new{Without clearly linking IT and OT systems in training, crews may face additional} operational risks, as a breach in IT systems could cascade into OT disruptions, potentially jeopardizing vessel safety and functionality. This finding underscores the need to bridge the IT-OT divide in cybersecurity training, as modern ships operate within interconnected systems that span both domains.\\
\noindent\textbf{\blackcircled{C} Preparedness for Real-World Cyber-Attacks.}
Participants were asked whether their cybersecurity training adequately prepared them for real-world incidents. The responses indicated \new{notable} gaps. As one participant observed, \emph{“[The training] didn’t even really say how to identify a cyber-attack; it talks a lot about flash drives”} (P9). Mariners reported learning about these attacks through direct experiences rather than structured training.
Another participant underscored the importance of traditional navigation skills in mitigating these gaps: \emph{“I’ve been trained in celestial navigation and charting. I know why it matters. It’s good seamanship. That’s what we rely on”} (P18). Enhancing training to include both cyber-specific threats and their operational consequences could better prepare mariners to identify and respond to such incidents. \\
\noindent\textbf{\blackcircled{D} \new{Challenges in} Prevention Practices.}
In response to questions about their day-to-day cybersecurity practices, participants highlighted challenges in balancing operational efficiency with adherence to cybersecurity protocols. Eight participants cited issues with practices like password changes, which interfere with task performance. Mariners frequently switch ships and spend long periods at home between assignments, leading to the use of insecure practices such as storing passwords on notes near computers. \emph{“You have to have logins to your computers… new guys are coming on board and all that. So a lot of times people will print it out, tape it to the bottom of the keyboard…”} (P5).

Additional responses revealed that some crew members bypassed security measures like locking USB ports due to a lack of understanding. \emph{“We were told to lock USB ports on computers, but some crew still found ways to bypass it because they didn’t fully understand the importance of this”} (P19). These findings underscore the need for practical, comprehensible protocols tailored to the realities of maritime operations. \\
\noindent\textbf{\blackcircled{E} Management of High-Bandwidth Satellite Internet.} The management of high-bandwidth satellite internet systems (e.g.,~Starlink) frequently emerged during discussions of shipboard cybersecurity and new technologies. The introduction of these systems aboard ships has provided faster internet access but has also introduced potential cybersecurity concerns due to decentralized management practices. Nine mariners described this decentralized management as a significant concern. \emph{“It’s the Wild West mostly. Every company is different. It’s not unified. And I gotta say probably most companies aren’t watching [the system] very closely”} (P4). A participant further highlighted these risks by referencing a case where a Navy Chief was demoted for improperly installing a Starlink system, illustrating how unregulated practices can lead to significant consequences~\cite{military2024navy}. The frequent mention of this issue suggests a pressing need for standardized policies governing emerging technologies like high-bandwidth satellite internet aboard ships.\\
\noindent\textbf{\blackcircled{F} Practices of Younger vs. Older Mariners.} In discussions of their confidence in handling cyber-attacks and critical equipment malfunctions, generational differences in navigation practices emerged. Five participants highlighted that younger mariners heavily rely on electronic systems such as ECDIS and GPS, often at the expense of traditional navigation skills. This dependence raises concerns about their ability to handle situations where all navigation electronics are compromised. One participant stated, \emph{“You just turn it off. Keep going. But, that’s also the old school. The new school guys don’t know how to drive boats without computers”} (P6). Another participant noted, \emph{“The younger generation relies entirely on GPS and ECDIS. They don’t know traditional navigation methods”} (P18). 

This reliance is exacerbated by the maritime industry’s transition to fully electronic navigation, with paper charts becoming obsolete. Four participants expressed concerns about losing paper charts, emphasizing the challenges of relying solely on electronic systems. As P13 explained, “A lot of ships are transitioning to paperless navigation… A big concern… is how redundant and resilient are the electronic systems on board.” Ironically, while younger mariners might be expected to excel in cyber-awareness (e.g., identifying phishing emails), their reliance on electronic systems creates significant gaps in their ability to respond to equipment malfunctions or cyber-attacks. With many ships no longer allowed to carry paper charts, this shift underscores a growing vulnerability in maritime operations increasingly reliant on electronic systems alone. \\
\noindent\textbf{\blackcircled{G} Education Impact on Cybersecurity Practices.}
When asked about their ability to respond to cyber-attacks, participants often reflected on how their educational backgrounds influenced their practices. The maritime industry allows individuals with diverse educational backgrounds to ascend to high-ranking positions, including captain, creating variability in cybersecurity awareness and response capabilities. As one participant noted, \emph{“You can make it all the way to captain in this industry, quarter million dollars a year, and have never graduated high school”} (P5). This variability can result in inconsistent handling of cyber events. Another participant highlighted gaps in basic technical understanding, stating, \emph{“I’d get woken up at like 8 P.M. because alarms are going off, and no one knows what the alarm is. Now you add on cyber security. They’re not being paid to know that”} (P2). This disparity underscores the need for standardized, accessible cybersecurity education tailored to mariners’ diverse backgrounds. \\
\noindent\textbf{\blackcircled{H} \new{Limited} Cyber-Attack Response Plans.}
Participants were asked whether their vessels had specific cyber-attack response plans. In total, 14 participants reported the absence of response plans for handling cyber-attacks, with many relying on vague instructions to “call IT.” As one mariner stated, \emph{“There’s no response plan or anything… If I’m asleep, there’s no plan, and 12 hours a day I’m not available”} (P2). This lack of preparedness often leaves crew members isolated during cyber incidents. Another participant highlighted the broader issue, explaining, \emph{“The crew on board is alone in case of a cyber-attack… IT guys are also not aware of the cyber threats because they don’t know the actual vessels… they are only familiar with the business network of the ship”} (P3). Some mariners resorted to makeshift responses, such as disconnecting Ethernet cords to contain potential threats, which one participant described as \emph{“a very rudimentary way of handling that, but quick”} (P9). These findings emphasize the need for comprehensive, vessel-specific cyber-attack response plans that address IT and OT challenges. \\
\noindent\textbf{\blackcircled{I} Cyber-Attack Response Confidence.}
Participants were asked about their confidence in dealing with cyber-attacks on ship. Mariners with real-world experience of cyber-attacks often expressed lower confidence in their ability to manage such incidents compared to those without firsthand experience. Among the 10 participants who had experienced cyber-attacks, 8 admitted to lacking confidence in handling these situations. For instance, one participant noted, \emph{“The ECDIS we have, you can’t even put in the position manually to fix what it’s saying after spoofing. It says I’m by Sicily, but I’m all the way by Cyprus”} (P18), reflecting a deep understanding of the limitations of their systems. Conversely, of the 11 participants who felt confident or unconcerned about handling cyber-attacks, only 2 had faced one in real life. This overconfidence often stemmed from misconceptions about the robustness of their systems, such as assuming that critical shipboard systems were inherently protected from external threats. These findings underscore the importance of targeted training to align mariners’ perceived and actual preparedness for managing cyber-attacks.

\begin{tcolorbox}[insightbox]
\textbf{\new{Basic} cybersecurity training, which overlooks \new{IT-OT interplay}, coupled with a lack of vessel-specific response plans, \new{may not fully equip mariners for} real-world threats like GPS spoofing and ransomware, \new{contributing to inconsistencies in} practices and\new{, in some cases,} overconfidence.}
\end{tcolorbox}

\subsection{Mariner Experiences with Regulation}
\label{sec:regulations}
This subsection explores mariners’ perspectives on maritime cybersecurity regulations, contrasting their views on traditional safety standards with the emerging cybersecurity requirements. By examining both general and cybersecurity-specific regulations, we aim to highlight differences in mariners’ perceptions and identify areas for improvement. Insights for this subsection were derived from the regulation and standards questions, which asked participants about their perceptions of current safety and security regulations, their thoughts on emerging cybersecurity standards, and how these frameworks impact their ability to maintain security on board.

\noindent\textbf{\blackcircled{J} Perceived Importance of Safety and Security Regulation.} 
To understand how mariners perceive the role of regulations in their work, we asked participants about their views on safety and security standards. This line of questioning aimed to assess whether mariners value such regulations and to identify any potential gaps or challenges in their implementation. Mariners generally viewed safety and security standards as essential and protective, with 15 participants mentioning this specifically. One participant emphasized this perspective, stating, \emph{“Most of the standards, regulations that we have in the maritime industry… are written in blood and oil. Mariners pay for it with their blood, sweat, and tears”} (P7). This sentiment reflects the deeply personal stakes mariners associate with regulatory frameworks designed to safeguard their well-being. \\
\noindent\textbf{\blackcircled{K} Drawbacks of Safety and Security Regulation.} 
We asked participants to describe their experiences with safety and security standards and whether they've experienced any challenges or drawbacks when following them. Responses to this question revealed several challenges, with 7 participants indicating that regulations often function as “catch-alls” and fail to account for ship-specific contexts or operational realities. One mariner explained, \emph{“A lot of retired Navy guys… they’re writing the regulation. But they’re writing it based off the Navy way of doing stuff with a 5,000 person crew on [an] aircraft carrier versus a 20 or 12 person crew”} (P6). Another noted, \emph{“I really think you have to tailor security to your operation, and I don’t think the regulations do that very well”} (P14).

Furthermore, 11 participants also expressed frustration with the burden these regulations impose, particularly as automation increases and crew sizes decrease. One mariner remarked, \emph{“As the engine room and other things become more automated, shipping companies are pushing for less mariners on board, but increasing the regulatory burden”} (P4). This additional workload can strain crews and impact compliance.

Mariners further noted that in certain situations, adhering to regulations might even be unsafe. For example, one participant shared, \emph{“There are times we have to bend the rules a little bit for security reasons. If you’re transiting off of Africa, don’t turn your lights on… So you’re not a target”} (P6). Others echoed this sentiment, emphasizing that strict adherence is not always practical or safe.

Finally, mariners criticized the reactive nature of the maritime industry regarding regulations. As one mariner put it, \emph{“The maritime industry is a very reactive industry… the standards get updated after [disasters] to cover new topics”} (P15). This reactive approach often leaves crews unprepared for emerging challenges and risks. \\
\noindent\textbf{\blackcircled{L} \new{Challenges in} Cybersecurity Rules and Regulation.} 
Participants were asked about their familiarity with and perceptions of emerging cybersecurity regulations, such as those from the International Maritime Organization (IMO). The IMO is a specialized United Nations agency responsible for global maritime safety and security standards, making its regulations particularly relevant for mariners. These questions aimed to uncover how mariners view the adequacy and implementation of cybersecurity standards. Of the 21 participants, only 10 were aware of the IMO's cybersecurity standards, and those who were aware often criticized them as impractical and unhelpful. Beyond the IMO, participants also expressed frustrations with general cybersecurity requirements and practices. For example, one mariner stated, \emph{“Password management is the bane of my existence”} (P4), reflecting the widespread challenge of implementing effective password policies across the maritime industry. Another participant \new{noted impractical training materials}, \emph{“There’s one PowerPoint we all go through, and it’s like don’t put flash drives on company computers… But then we order flash drives because there’s a lot of computers, a lot of things gotta get done”} (P9).

Six participants expressed frustration at being held responsible for cybersecurity without proper training. As one put it, \emph{“I already have to deal with so much. I don’t want to be liable for something I’m not trained to do”} (P2). This lack of expertise led four participants to advocate for the inclusion of a dedicated cybersecurity professional on board. One participant explained, \emph{“We’re at the point with networking where you need somebody with more knowledge than what we currently have on board”} (P7). Another added, \emph{“You need a full-time cybersecurity person on board for an operation like us… especially with MSC, these bigger crews”} (P9).

The absence of cybersecurity standards for older vessels was also a point of contention. One mariner noted, \emph{“IAX standards also should improve the cyber security standards of the shipping industry. Currently, they publish 2 guidelines… for new constructed ships, not old vessels”} (P3). IAX refers to specific IMO guidelines aimed at enhancing cybersecurity on ships.

Five participants further expressed concerns about the loss of paper navigation charts due to new regulations, which they felt could increase vulnerability. One mariner explained, \emph{“If you get an EMP pulse and it fries all [your navigation equipment], I don’t know how the ship would ever go to sea without a paper chart”} (P10). Another commented, \emph{“If you don’t have charts and you’re being spoofed, you’re a little screwed”} (P18). These perspectives highlight the need for more practical and inclusive cybersecurity standards that address the realities of modern maritime operations.

\begin{tcolorbox}[insightbox]
\textbf{Mariners recognize the importance of safety and security regulations but find them burdensome and poorly aligned with operational realities. Cybersecurity rules, in particular, are seen as reactive, impractical, and insufficiently tailored to address training gaps, older vessels, and emerging threats.}
\end{tcolorbox}

\section{Concluding Discussion}
This work highlights critical gaps in maritime cybersecurity, emphasizing the need for tailored solutions to address the unique challenges faced by mariners and ship systems. From the disconnect between cybersecurity training and real-world threats, to the practical limitations of current regulations, our findings underscore the importance of industry-wide improvements. Mariners’ experiences and perspectives reveal the pressing need for enhanced training, robust security frameworks, and better integration of technology with practical operations.  These results not only identify areas for immediate action but also lay the groundwork for future research. The following subsections provide actionable recommendations to enhance training, improve detection and response capabilities, and refine regulatory frameworks, along with outlining open research directions to address emerging challenges in this critical domain.

\begin{figure}[t!]
    \centering  
    \includegraphics[width=\linewidth]{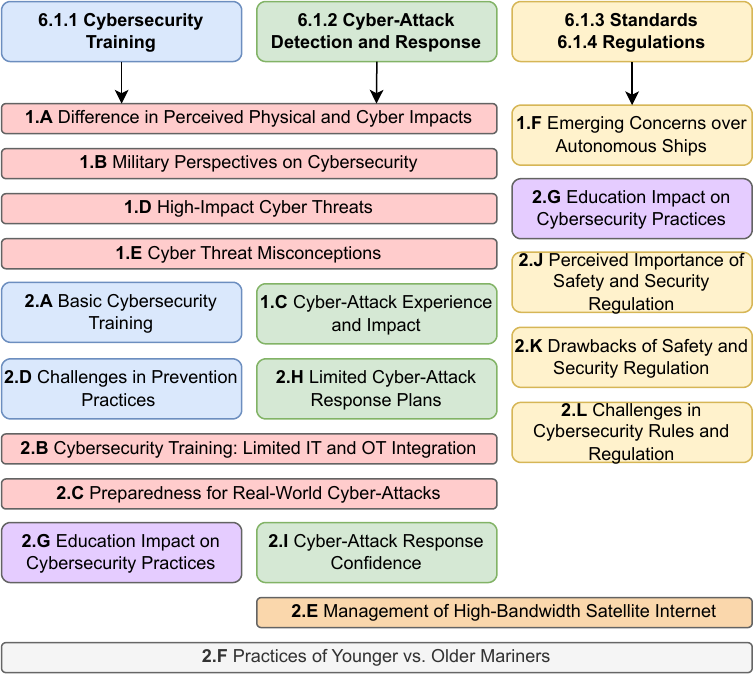}
    \caption{\new{Map of Results to Discussion Recommendations}}
    \label{fig:insights}
\end{figure}

\subsection{Recommendations}
Here, we first present actionable recommendations grounded in the insights from mariner interviews. As illustrated in Figure~\ref{fig:insights}, the recommendations align with key findings from the results sections, with each insight tied to specific recommendations to ensure relevance. This serves as a holistic guide to understanding the connections between the results and the proposed improvements. Grounded in mariners’ experiences and identified gaps in training, practices, and regulations, these recommendations address critical challenges in maritime cybersecurity.

\subsubsection{Enhancing Cybersecurity Training}
Improving mariner cybersecurity preparedness starts with addressing the gaps in training identified during the study. Effective training must provide role-specific and operationally relevant content that translates into practical, onboard skills. \new{ICS research has shown that training often fails when not designed around end-user roles and workflows~\cite{nunes2024exploiting,fla2024challenges}.}

The diversity of mariners’ educational and experiential backgrounds adds unique challenges to designing effective cybersecurity training. Mariners range from high school graduates advancing through the ranks to maritime academy degree holders. Their exposure to ship types also varies; some gain broad experience across vessels, while others remain on a single type, limiting their operational knowledge~(\Cref{sec:training_practices} –- G). Unlike aviation or power generation, where standardized training is the norm, this diversity demands flexible cybersecurity training tailored to varied backgrounds, ensuring all mariners can navigate the unique challenges of maritime cybersecurity. \new{This aligns with research showing that OT personnel's mindsets, shaped by operational demands and safety culture, benefit most from practical, context-specific security training tailored to their real-world tasks~\cite{evripidou2022understanding}.}

To compound these challenges, the results revealed significant gaps in mariner cybersecurity training, with participants describing it as overly simplistic and disconnected from operational realities~(\Cref{sec:training_practices} –- A). \new{Similarly, a recent study of phishing training in high-risk sectors found that generic modules were often ignored or rushed through, limiting their practical value~\cite{ho2025phishing}. These findings underscore the need to move away from impersonal methods such as PowerPoints or computer modules. Instead, training should be delivered through in-person, hands-on approaches that actively engage mariners.} This tailored approach should address ship-specific contexts and mariners’ unique responsibilities onboard, making training more relevant and practical~(\Cref{sec:training_practices} –- C). For example, the disconnect between IT and OT systems must be addressed by incorporating content that highlights how the cybersecurity of a ship’s computing assets directly impacts its physical operational safety~(\Cref{sec:training_practices} –- B).

The findings also highlight insufficient education on the integration of control engineering and cybersecurity, as well as a misunderstanding of the dynamic interconnection between cyber and physical systems. Mariners often \new{do not} associate cyber risks with their potential physical consequences, such as financial loss, operational disruption, or endangerment of lives~(\Cref{sec:mental_model} –- A). Training programs must explicitly bridge this gap by illustrating how cyberattacks, such as ransomware or system sabotage, can have direct, tangible impacts on maritime operations. For example, emphasizing scenarios where a compromised GPS could lead to collisions or grounding would make the connection between cybersecurity and physical security more explicit. Addressing this disconnect through domain-specific educational initiatives would reduce risky behaviors, such as poor password management or bypassing USB restrictions, while encouraging mariners to adopt better cybersecurity practices onboard~(\Cref{sec:training_practices} –- D). \new{Sustaining such improvements also requires reinforcing secure actions through feasible, habit-building routines~\cite{sasse2022rebooting}.}

Training content should be customized based on the mariners’ roles onboard. For instance, officers responsible for navigation would benefit from training that emphasizes detecting and responding to GPS or AIS spoofing, while engineers managing \cam{OT} systems should focus on identifying vulnerabilities in remotely monitored systems and mitigating their exploitation. Similarly, crew members working with communication systems might require a deeper understanding of email phishing and network vulnerabilities~(\Cref{sec:mental_model} –- D). By aligning the curriculum with the distinct operational roles on a ship, each mariner can develop the specialized competencies needed to effectively mitigate cybersecurity risks, thereby enhancing the overall security posture of maritime operations. \new{However, it is equally important to recognize that adding new training duties can be burdensome for mariners already handling demanding workloads, especially in smaller companies with fewer resources for specialized staff or frequent drills. Larger organizations may be able to offset this through onboard cybersecurity specialists or dedicated trainers, while smaller companies might require cost-effective support, such as remote consultants or shared training resources. Recent work in OT settings shows that external consultants can support training efforts, but their impact depends on whether internal teams can sustain the changes~\cite{evripidou2023exploring}. \cam{Complementary improvements in usability and system design can help embed these changes into daily practice.}}

\subsubsection{Improving Cyber-Attack Detection and Response}
Bridging the gap between knowledge and operational readiness is key to enhancing mariners’ ability to respond to cyber-attacks. These recommendations focus on equipping mariners with practical, actionable measures for detecting and mitigating real-time threats.

A significant gap identified in this study is the absence of robust, actionable cyber-attack response plans~(\Cref{sec:training_practices} -- H, I). To address this gap, response plans must extend beyond the current “call IT” standard, incorporating onboard protocols accessible to mariners in critical situations. For example, common cyber threats such as GPS spoofing, AIS jamming, and radar manipulation should be central to training and response protocols~(\Cref{sec:mental_model} -- C). \new{This is consistent with a recent usability study, which shows that giving operators clear, step-by-step guidance for security features can boost confidence in responding to critical threats~\cite{li2024windows98}.} \new{Some of this burden could be reduced through automation or remote assistance from shore-based IT teams, though connectivity, bandwidth, and resource limitations may hinder their reliability in practice.}

Furthermore, hands-on response training can prepare mariners to detect attacks more effectively. For instance, correcting misconceptions about system vulnerabilities and teaching the interconnected nature of OT and IT systems would enhance awareness~(\Cref{sec:mental_model} –- E). The importance of network segregation, especially regarding user-installed systems like Starlink, must also be emphasized~(\Cref{sec:training_practices} –- E). One participant recounted the story of a Navy Chief being demoted for improper Starlink installation~(\Cref{sec:training_practices} –- E), which starkly contrasts with the informal practices observed in the industry and underscores the need for clear policies and training~\cite{military2024navy}.

Adopting practices from the \cam{MSC}, such as MCON drills that build confidence in navigating without electronic systems~(\Cref{sec:mental_model} -- B; \Cref{sec:training_practices} –- F), could serve as a model for broader industry adoption. Mariners trained in these methods consistently reported higher confidence and readiness in handling cyber-attack scenarios.

\subsubsection{Toward Domain-Specific Cybersecurity Standards}
Adapting industry regulations to align with the realities of maritime operations is crucial. As highlighted in the findings, existing regulations often fail to account for ship-specific contexts and operational constraints, leaving mariners frustrated and compliance challenging~(\Cref{sec:regulations} –- J, K). For example, participants described safety and security regulations as overly generic, designed for large-scale operations but poorly adapted to smaller crews and merchant vessels. Addressing these gaps requires collaboration with mariners to identify and amend impractical or overly burdensome rules. \new{This mirrors findings from the energy sector, where interdisciplinary collaboration between cybersecurity and OT personnel has been shown to improve adoption of standards by aligning them with practical, operational constraints~\cite{gallardo2024interdisciplinary}.}

Emerging technologies, including Starlink and remotely monitored equipment, introduce additional complexities. Many participants expressed concerns over the decentralized management of these systems and the lack of standardized oversight, emphasizing the need for mandatory penetration testing and secure integration~(\Cref{sec:training_practices} –- E). Similarly, the shift toward autonomous and remotely operated ships increases reliance on interconnected systems, amplifying cybersecurity risks~(\Cref{sec:mental_model} –- F). The findings reveal the maritime industry’s reactive regulatory culture, where standards are developed post-incident rather than proactively addressing emerging threats~(\Cref{sec:regulations} –- K). Participants noted that the transition to fully electronic navigation without sufficient redundancy exacerbates vulnerabilities, highlighting the need for practical, ship-specific regulations that reflect operational realities.

Finally, the secretive nature of the industry around cyber-attack disclosures hinders progress. Establishing anonymous reporting mechanisms for cyber incidents could provide regulators with critical insights to develop more effective policies~(\Cref{sec:regulations} –- L). Programs like bug bounties for ship equipment could also incentivize innovation and accountability while improving system security.

\subsubsection{Unified Cybersecurity Regulation for the Maritime Sector}
The maritime sector’s resilience against cyber threats depends on a globally unified regulatory framework for maritime cybersecurity~(\Cref{sec:regulations} -- L). The Department of Homeland Security (DHS) identifies 16 critical infrastructure sectors vital to the nation’s security, economy, and public health, including the Maritime Transportation System (MTS) within the Transportation Systems Sector \cite{us_dhs_infrastructure}. Despite its critical role, the maritime sector’s cybersecurity regulations are fragmented compared to robust frameworks like the NERC Critical Infrastructure Protection (NERC CIP) standards applied in the Energy Sector~\cite{nerc_cip}. NERC CIP mandates detailed, enforceable requirements across all critical energy operators, ensuring consistency and security in addressing cyber threats. These include specific controls for access management, incident reporting, and system resilience. In contrast, maritime cybersecurity regulations, such as IMO Resolution MSC.428(98) and BIMCO Guidelines, focus on high-level risk management and lack the specificity and enforceability seen in NERC CIP \cite{cyber_guidelines_v5,imo_msc428}.

To address these gaps, maritime cybersecurity regulations should adopt a unified framework inspired by NERC CIP, which includes:
\begin{itemize}[nosep, leftmargin=*]
    \item \textbf{Standardizing Practices Globally}: Creating detailed, enforceable requirements applicable across maritime operators, regardless of size or geography.
    \item \textbf{Enhancing Specificity}: Moving beyond high-level risk assessments to include prescriptive technical controls, such as network segmentation, encryption, and access management.
    \item \textbf{Mandating Incident Reporting}: Introducing mandatory, anonymous incident disclosure mechanisms to improve the industry’s understanding of threats and vulnerabilities.
    \item \textbf{Addressing Emerging Technologies}: Developing regulations specific to new systems like Starlink, remotely monitored engines, and other Internet-connected shipboard equipment.
    \item \textbf{Incentivizing Compliance}: Providing financial or operational incentives for operators to meet these enhanced standards, similar to the energy sector’s approach.
\end{itemize}

By aligning maritime cybersecurity regulations with the rigor of frameworks like NERC CIP, the maritime sector can strengthen its resilience against cyber threats. A unified, enforceable approach would simplify compliance for operators while ensuring best practices are consistently applied across the global supply chain.

\subsection{Future Research Directions}
This study has uncovered significant gaps in maritime cybersecurity, which can inform future research efforts. Below are potential directions for future work, each addressing critical issues revealed in the user study results. \\
\noindent\textbf{Real-Time Detection for Maritime Cyber-Physical Systems.} Developing real-time intrusion detection systems (IDS) for maritime cyber-physical systems is essential for enhancing security. Drawing parallels with automotive security, these systems must be tailored to ship-specific protocols like NMEA 2000 to address unique vulnerabilities and better protect shipboard networks. Future work could focus on lightweight, real-time solutions optimized for maritime-specific challenges, such as remote locations and satellite internet connectivity. This approach would empower crews to respond effectively, minimizing disruptions to maritime operations. \\
\noindent\textbf{Proactive Risk Assessment of Shipboard Systems.} Comprehensive security risk assessments of maritime cyber-physical systems are essential to identifying and mitigating vulnerabilities in IT and OT components. Unlike real-time detection systems, this approach emphasizes proactive measures to strengthen defenses and prevent attacks. Research could explore risks in interconnected systems like navigation, engine monitoring, and onboard communication networks, leading to targeted hardening techniques and best practices. \\
\noindent\textbf{Qualitative Research in Maritime Cybersecurity.} Future qualitative research could focus on specific maritime groups identified in this study for their unique perspectives. Two participants involved in deep-sea operations and nine military-affiliated mariners offered distinct insights into cybersecurity challenges. Deep-sea operators manage critical infrastructure like subsea cables, vital for global internet connectivity and highly vulnerable to cyber threats. Military-affiliated mariners bring specialized training and protocols, such as those informed by \cam{MSC} practices, which could serve as benchmarks for broader industry improvements. Researching these groups further could reveal valuable strategies for managing cybersecurity in diverse and critical operational contexts.

\section{Acknowledgments}
We thank the anonymous reviewers for their constructive feedback. We also thank the mariners who generously shared their time and experiences; their insights were indispensable to this study. This work was funded in part by the National Science Foundation (NSF).

\bibliographystyle{ACM-Reference-Format.bst}
\bibliography{refs}


\begin{thebibliography}{115}


\ifx \showCODEN    \undefined \def \showCODEN     #1{\unskip}     \fi
\ifx \showDOI      \undefined \def \showDOI       #1{#1}\fi
\ifx \showISBNx    \undefined \def \showISBNx     #1{\unskip}     \fi
\ifx \showISBNxiii \undefined \def \showISBNxiii  #1{\unskip}     \fi
\ifx \showISSN     \undefined \def \showISSN      #1{\unskip}     \fi
\ifx \showLCCN     \undefined \def \showLCCN      #1{\unskip}     \fi
\ifx \shownote     \undefined \def \shownote      #1{#1}          \fi
\ifx \showarticletitle \undefined \def \showarticletitle #1{#1}   \fi
\ifx \showURL      \undefined \def \showURL       {\relax}        \fi
\providecommand\bibfield[2]{#2}
\providecommand\bibinfo[2]{#2}
\providecommand\natexlab[1]{#1}
\providecommand\showeprint[2][]{arXiv:#2}

\bibitem[UNCTAD(2024)]%
        {unctad_maritime_transport_2024}
\bibfield{author}{\bibinfo{person}{UNCTAD}.} \bibinfo{year}{2024}\natexlab{}.
\newblock \bibinfo{title}{Launch of the Review of Maritime Transport 2024}.
\newblock
\newblock
\newblock
\shownote{\href{https://unctad.org/meeting/launch-review-maritime-transport-2024}{unctad.org}}.


\bibitem[Kottasová(2024)]%
        {cnn2024subseacables}
\bibfield{author}{\bibinfo{person}{Ivana Kottasová}.}
  \bibinfo{year}{2024}\natexlab{}.
\newblock \showarticletitle{Sabotage suspected as undersea cables cut in the
  Baltic Sea}.
\newblock \bibinfo{journal}{\emph{CNN}} (\bibinfo{year}{2024}).
\newblock
\newblock
\shownote{\href{https://www.cnn.com/2024/11/19/europe/sabotage-undersea-cables-cut-baltic-sea-intl/index.html}{cnn.com}}.


\bibitem[Guardian(2024)]%
        {guardian2024underseacables}
\bibfield{author}{\bibinfo{person}{The Guardian}.}
  \bibinfo{year}{2024}\natexlab{}.
\newblock \showarticletitle{Sweden seeks clarity from China about suspected
  sabotage of undersea cables}.
\newblock \bibinfo{journal}{\emph{The Guardian}} (\bibinfo{year}{2024}).
\newblock
\newblock
\shownote{\href{https://www.theguardian.com/world/2024/nov/28/sweden-seeks-clarity-from-china-about-suspected-sabotage-of-undersea-cables}{theguardian.com}}.


\bibitem[Chappell(2024)]%
        {npr2024baltimore}
\bibfield{author}{\bibinfo{person}{Bill Chappell}.}
  \bibinfo{year}{2024}\natexlab{}.
\newblock \showarticletitle{U.S. sues Dali ship owner and operator for \$100
  million over Baltimore bridge collapse}.
\newblock \bibinfo{journal}{\emph{NPR}} (\bibinfo{year}{2024}).
\newblock
\newblock
\shownote{\href{https://www.npr.org/2024/09/18/nx-s1-5117681/us-justice-suit-baltimore-key-bridge-collapse-dali-ship}{npr.org}}.


\bibitem[Skene(2024)]%
        {ap2024baltimore}
\bibfield{author}{\bibinfo{person}{Lea Skene}.}
  \bibinfo{year}{2024}\natexlab{}.
\newblock \showarticletitle{Baltimore bridge collapses after powerless cargo
  ship rams into support column; 6 presumed dead}.
\newblock \bibinfo{journal}{\emph{AP News}} (\bibinfo{year}{2024}).
\newblock
\newblock
\shownote{\href{https://apnews.com/article/baltimore-bridge-collapse-53169b379820032f832de4016c655d1b}{apnews.com}}.


\bibitem[News(2021)]%
        {bbc2021suez}
\bibfield{author}{\bibinfo{person}{BBC News}.} \bibinfo{year}{2021}\natexlab{}.
\newblock \showarticletitle{Egypt's Suez Canal blocked by huge container ship}.
\newblock \bibinfo{journal}{\emph{BBC News}} (\bibinfo{year}{2021}).
\newblock
\newblock
\shownote{\href{https://www.bbc.com/news/world-middle-east-56505413}{bbc.com}}.


\bibitem[Das(2021)]%
        {das2021suez}
\bibfield{author}{\bibinfo{person}{Koustav Das}.}
  \bibinfo{year}{2021}\natexlab{}.
\newblock \showarticletitle{Explained: How much did Suez Canal blockage cost
  world trade}.
\newblock \bibinfo{journal}{\emph{India Today}} (\bibinfo{year}{2021}).
\newblock
\newblock
\shownote{\href{https://www.indiatoday.in/business/story/explained-how-much-did-suez-canal-blockage-cost-world-trade-1785062-2021-03-30}{indiatoday.in}}.


\bibitem[Greig(2023)]%
        {greig2024ransomware}
\bibfield{author}{\bibinfo{person}{Jonathan Greig}.}
  \bibinfo{year}{2023}\natexlab{}.
\newblock \showarticletitle{Ransomware attack on maritime software impacts
  1,000 ships}.
\newblock \bibinfo{journal}{\emph{The Record}} (\bibinfo{year}{2023}).
\newblock
\newblock
\shownote{\href{https://therecord.media/ransomware-attack-on-maritime-software-impacts-1000-ships}{therecord.media}}.


\bibitem[DNV(2024)]%
        {dnv_maritime_cybersecurity_2024}
\bibfield{author}{\bibinfo{person}{DNV}.} \bibinfo{year}{2024}\natexlab{}.
\newblock \bibinfo{title}{Maritime cyber security: Mandatory IACS Unified
  Requirements for newbuilds from July 2024}.
\newblock
\newblock
\newblock
\shownote{\href{https://www.dnv.com/maritime/webinars-and-videos/on-demand-webinars/access/maritime-cyber-security-mandatory-iacs-ur-for-newbuilds-from-July2024-23may/}{dnv.com}}.


\bibitem[Pavur et~al\mbox{.}(2020)]%
        {pavur2020tale}
\bibfield{author}{\bibinfo{person}{James Pavur}, \bibinfo{person}{Daniel
  Moser}, \bibinfo{person}{Martin Strohmeier}, \bibinfo{person}{Vincent
  Lenders}, {and} \bibinfo{person}{Ivan Martinovic}.}
  \bibinfo{year}{2020}\natexlab{}.
\newblock \showarticletitle{A Tale of Sea and Sky On the Security of Maritime
  VSAT Communications}. In \bibinfo{booktitle}{\emph{IEEE S\&P}}.
\newblock


\bibitem[Liu et~al\mbox{.}(2021)]%
        {liu2021stars}
\bibfield{author}{\bibinfo{person}{Shinan Liu}, \bibinfo{person}{Xiang Cheng},
  \bibinfo{person}{Hanchao Yang}, \bibinfo{person}{Yuanchao Shu},
  \bibinfo{person}{Xiaoran Weng}, \bibinfo{person}{Ping Guo},
  \bibinfo{person}{Kexiong~Curtis Zeng}, \bibinfo{person}{Gang Wang}, {and}
  \bibinfo{person}{Yaling Yang}.} \bibinfo{year}{2021}\natexlab{}.
\newblock \showarticletitle{Stars can tell: a robust method to defend against
  GPS spoofing attacks using off-the-shelf chipset}. In
  \bibinfo{booktitle}{\emph{USENIX Security}}.
\newblock


\bibitem[Lundberg et~al\mbox{.}(2014)]%
        {lundberg2014security}
\bibfield{author}{\bibinfo{person}{Devin Lundberg}, \bibinfo{person}{Brown
  Farinholt}, \bibinfo{person}{Edward Sullivan}, \bibinfo{person}{Ryan Mast},
  \bibinfo{person}{Stephen Checkoway}, \bibinfo{person}{Stefan Savage},
  \bibinfo{person}{Alex~C Snoeren}, {and} \bibinfo{person}{Kirill Levchenko}.}
  \bibinfo{year}{2014}\natexlab{}.
\newblock \showarticletitle{On the security of mobile cockpit information
  systems}. In \bibinfo{booktitle}{\emph{CCS}}.
\newblock


\bibitem[Birnbach et~al\mbox{.}(2017)]%
        {birnbach2017wi}
\bibfield{author}{\bibinfo{person}{Simon Birnbach}, \bibinfo{person}{Richard
  Baker}, {and} \bibinfo{person}{Ivan Martinovic}.}
  \bibinfo{year}{2017}\natexlab{}.
\newblock \showarticletitle{Wi-fly?: Detecting privacy invasion attacks by
  consumer drones}. In \bibinfo{booktitle}{\emph{NDSS}}.
\newblock


\bibitem[Jansen et~al\mbox{.}(2017)]%
        {jansen2017localization}
\bibfield{author}{\bibinfo{person}{Kai Jansen}, \bibinfo{person}{Matthias
  Sch{\"a}fer}, \bibinfo{person}{Vincent Lenders}, \bibinfo{person}{Christina
  P{\"o}pper}, {and} \bibinfo{person}{Jens Schmitt}.}
  \bibinfo{year}{2017}\natexlab{}.
\newblock \showarticletitle{Localization of spoofing devices using a
  large-scale air traffic surveillance system}. In
  \bibinfo{booktitle}{\emph{ASIACCS}}.
\newblock


\bibitem[Jansen et~al\mbox{.}(2021)]%
        {jansen2021trust}
\bibfield{author}{\bibinfo{person}{Kai Jansen}, \bibinfo{person}{Liang Niu},
  \bibinfo{person}{Nian Xue}, \bibinfo{person}{Ivan Martinovic}, {and}
  \bibinfo{person}{Christina P{\"o}pper}.} \bibinfo{year}{2021}\natexlab{}.
\newblock \showarticletitle{Trust the Crowd: Wireless Witnessing to Detect
  Attacks on ADS-B-Based Air-Traffic Surveillance}. In
  \bibinfo{booktitle}{\emph{NDSS}}.
\newblock


\bibitem[Longo et~al\mbox{.}(2024)]%
        {longo2024collision}
\bibfield{author}{\bibinfo{person}{Giacomo Longo}, \bibinfo{person}{Martin
  Strohmeier}, \bibinfo{person}{Enrico Russo}, \bibinfo{person}{Alessio Merlo},
  {and} \bibinfo{person}{Vincent Lenders}.} \bibinfo{year}{2024}\natexlab{}.
\newblock \showarticletitle{On a Collision Course: Unveiling Wireless Attacks
  to the Aircraft Traffic Collision Avoidance System (TCAS)}. In
  \bibinfo{booktitle}{\emph{USENIX Security}}.
\newblock


\bibitem[Rouf et~al\mbox{.}(2010)]%
        {rouf2010security}
\bibfield{author}{\bibinfo{person}{Ishtiaq Rouf}, \bibinfo{person}{Rob Miller},
  \bibinfo{person}{Hossen Mustafa}, \bibinfo{person}{Travis Taylor},
  \bibinfo{person}{Sangho Oh}, \bibinfo{person}{Wenyuan Xu},
  \bibinfo{person}{Marco Gruteser}, \bibinfo{person}{Wade Trappe}, {and}
  \bibinfo{person}{Ivan Seskar}.} \bibinfo{year}{2010}\natexlab{}.
\newblock \showarticletitle{Security and privacy vulnerabilities of In-Car
  wireless networks: A tire pressure monitoring system case study}. In
  \bibinfo{booktitle}{\emph{USENIX Security}}.
\newblock


\bibitem[Bhatia et~al\mbox{.}(2021)]%
        {bhatia2021evading}
\bibfield{author}{\bibinfo{person}{Rohit Bhatia}, \bibinfo{person}{Vireshwar
  Kumar}, \bibinfo{person}{Khaled Serag}, \bibinfo{person}{Z~Berkay Celik},
  \bibinfo{person}{Mathias Payer}, {and} \bibinfo{person}{Dongyan Xu}.}
  \bibinfo{year}{2021}\natexlab{}.
\newblock \showarticletitle{Evading Voltage-Based Intrusion Detection on
  Automotive CAN}. In \bibinfo{booktitle}{\emph{NDSS}}.
\newblock


\bibitem[Checkoway et~al\mbox{.}(2011)]%
        {checkoway2011comprehensive}
\bibfield{author}{\bibinfo{person}{Stephen Checkoway}, \bibinfo{person}{Damon
  McCoy}, \bibinfo{person}{Brian Kantor}, \bibinfo{person}{Danny Anderson},
  \bibinfo{person}{Hovav Shacham}, \bibinfo{person}{Stefan Savage},
  \bibinfo{person}{Karl Koscher}, \bibinfo{person}{Alexei Czeskis},
  \bibinfo{person}{Franziska Roesner}, {and} \bibinfo{person}{Tadayoshi
  Kohno}.} \bibinfo{year}{2011}\natexlab{}.
\newblock \showarticletitle{Comprehensive experimental analyses of automotive
  attack surfaces}. In \bibinfo{booktitle}{\emph{USENIX Security}}.
\newblock


\bibitem[Cho and Shin(2016)]%
        {cho2016fingerprinting}
\bibfield{author}{\bibinfo{person}{Kyong-Tak Cho} {and} \bibinfo{person}{Kang~G
  Shin}.} \bibinfo{year}{2016}\natexlab{}.
\newblock \showarticletitle{Fingerprinting electronic control units for vehicle
  intrusion detection}. In \bibinfo{booktitle}{\emph{USENIX Security}}.
\newblock


\bibitem[Garcia et~al\mbox{.}(2016)]%
        {garcia2016lock}
\bibfield{author}{\bibinfo{person}{Flavio~D Garcia}, \bibinfo{person}{David
  Oswald}, \bibinfo{person}{Timo Kasper}, {and} \bibinfo{person}{Pierre
  Pavlid{\`e}s}.} \bibinfo{year}{2016}\natexlab{}.
\newblock \showarticletitle{Lock it and still lose it—on the (In) Security of
  automotive remote keyless entry systems}. In \bibinfo{booktitle}{\emph{USENIX
  Security}}.
\newblock


\bibitem[Hu et~al\mbox{.}(2021)]%
        {hu2021automated}
\bibfield{author}{\bibinfo{person}{Shengtuo Hu}, \bibinfo{person}{Qi~Alfred
  Chen}, \bibinfo{person}{Jiachen Sun}, \bibinfo{person}{Yiheng Feng},
  \bibinfo{person}{Z~Morley Mao}, {and} \bibinfo{person}{Henry~X Liu}.}
  \bibinfo{year}{2021}\natexlab{}.
\newblock \showarticletitle{Automated Discovery of Denial-of-Service
  Vulnerabilities in Connected Vehicle Protocols}. In
  \bibinfo{booktitle}{\emph{USENIX Security}}.
\newblock


\bibitem[Jing et~al\mbox{.}(2024)]%
        {jing2024revisiting}
\bibfield{author}{\bibinfo{person}{Pengfei Jing}, \bibinfo{person}{Zhiqiang
  Cai}, \bibinfo{person}{Yingjie Cao}, \bibinfo{person}{Le Yu},
  \bibinfo{person}{Yuefeng Du}, \bibinfo{person}{Wenkai Zhang},
  \bibinfo{person}{Chenxiong Qian}, \bibinfo{person}{Xiapu Luo},
  \bibinfo{person}{Sen Nie}, {and} \bibinfo{person}{Shi Wu}.}
  \bibinfo{year}{2024}\natexlab{}.
\newblock \showarticletitle{Revisiting automotive attack surfaces: a
  practitioners’ perspective}. In \bibinfo{booktitle}{\emph{IEEE S\&P}}.
\newblock


\bibitem[Koscher et~al\mbox{.}(2010)]%
        {koscher2010experimental}
\bibfield{author}{\bibinfo{person}{Karl Koscher}, \bibinfo{person}{Alexei
  Czeskis}, \bibinfo{person}{Franziska Roesner}, \bibinfo{person}{Shwetak
  Patel}, \bibinfo{person}{Tadayoshi Kohno}, \bibinfo{person}{Stephen
  Checkoway}, \bibinfo{person}{Damon McCoy}, \bibinfo{person}{Brian Kantor},
  \bibinfo{person}{Danny Anderson}, \bibinfo{person}{Hovav Shacham},
  {et~al\mbox{.}}} \bibinfo{year}{2010}\natexlab{}.
\newblock \showarticletitle{Experimental security analysis of a modern
  automobile}. In \bibinfo{booktitle}{\emph{IEEE S\&P}}.
\newblock


\bibitem[Tang et~al\mbox{.}(2024)]%
        {tang2024eracan}
\bibfield{author}{\bibinfo{person}{Zhaozhou Tang}, \bibinfo{person}{Khaled
  Serag}, \bibinfo{person}{Saman Zonouz}, \bibinfo{person}{Z~Berkay Celik},
  \bibinfo{person}{Dongyan Xu}, {and} \bibinfo{person}{Raheem Beyah}.}
  \bibinfo{year}{2024}\natexlab{}.
\newblock \showarticletitle{ERACAN: Defending Against an Emerging CAN Threat
  Model}. In \bibinfo{booktitle}{\emph{CCS}}.
\newblock


\bibitem[Wen et~al\mbox{.}(2020)]%
        {wen2020plug}
\bibfield{author}{\bibinfo{person}{Haohuang Wen}, \bibinfo{person}{Qi~Alfred
  Chen}, {and} \bibinfo{person}{Zhiqiang Lin}.}
  \bibinfo{year}{2020}\natexlab{}.
\newblock \showarticletitle{Plug-N-Pwned: Comprehensive vulnerability analysis
  of OBD-II dongles as a new Over-the-Air attack surface in automotive IoT}. In
  \bibinfo{booktitle}{\emph{USENIX Security}}.
\newblock


\bibitem[Xue et~al\mbox{.}(2022)]%
        {xue2022said}
\bibfield{author}{\bibinfo{person}{Lei Xue}, \bibinfo{person}{Yangyang Liu},
  \bibinfo{person}{Tianqi Li}, \bibinfo{person}{Kaifa Zhao},
  \bibinfo{person}{Jianfeng Li}, \bibinfo{person}{Le Yu},
  \bibinfo{person}{Xiapu Luo}, \bibinfo{person}{Yajin Zhou}, {and}
  \bibinfo{person}{Guofei Gu}.} \bibinfo{year}{2022}\natexlab{}.
\newblock \showarticletitle{SAID: State-aware defense against injection attacks
  on in-vehicle network}. In \bibinfo{booktitle}{\emph{USENIX Security}}.
\newblock


\bibitem[Zhu et~al\mbox{.}(2024)]%
        {zhu2024ae}
\bibfield{author}{\bibinfo{person}{Shenchen Zhu}, \bibinfo{person}{Yue Zhao},
  \bibinfo{person}{Kai Chen}, \bibinfo{person}{Bo Wang},
  \bibinfo{person}{Hualong Ma}, {et~al\mbox{.}}}
  \bibinfo{year}{2024}\natexlab{}.
\newblock \showarticletitle{AE-Morpher: Improve Physical Robustness of
  Adversarial Objects against LiDAR-based Detectors via Object Reconstruction}.
  In \bibinfo{booktitle}{\emph{USENIX Security}}.
\newblock


\bibitem[Britannia(2024)]%
        {britannia}
\bibfield{author}{\bibinfo{person}{Britannia}.}
  \bibinfo{year}{2024}\natexlab{}.
\newblock \bibinfo{title}{Third Party Workers}.
\newblock
\newblock
\newblock
\shownote{\href{https://britanniapandi.com/2024/07/third-party-workers/}{britanniapandi.com}}.


\bibitem[maritimecyprus(2024)]%
        {maritimecyprus}
\bibfield{author}{\bibinfo{person}{maritimecyprus}.}
  \bibinfo{year}{2024}\natexlab{}.
\newblock \bibinfo{title}{Maritime Loss Prevention: Third Party Workers onboard
  the vessel}.
\newblock
\newblock
\newblock
\shownote{\href{https://maritimecyprus.com/2024/10/31/maritime-loss-prevention-third-party-workers-onboard-the-vessel/}{maritimecyprus.com}}.


\bibitem[Bryant(2012)]%
        {useless_twic}
\bibfield{author}{\bibinfo{person}{Dennis~L. Bryant}.}
  \bibinfo{year}{2012}\natexlab{}.
\newblock \bibinfo{title}{Maritime Security \& The Useless TWIC}.
\newblock
\newblock
\newblock
\shownote{\href{https://www.marinelink.com/news/maritime-security-useless344893}{marinelink.com}}.


\bibitem[Wynters(2023)]%
        {twic_threat}
\bibfield{author}{\bibinfo{person}{Lauren Wynters}.}
  \bibinfo{year}{2023}\natexlab{}.
\newblock \bibinfo{title}{The Growing Security Threat of Fake TWIC
  Credentials}.
\newblock
\newblock
\newblock
\shownote{\href{https://www.magnar.com/the-growing-security-threat-of-fake-twic-credentials/}{magnar.com}}.


\bibitem[DHS(2024)]%
        {us_dhs_cybersecurity_marine_2024}
\bibfield{author}{\bibinfo{person}{U.S. DHS}.} \bibinfo{year}{2024}\natexlab{}.
\newblock \bibinfo{title}{Cybersecurity in the MTS}.
\newblock
\newblock
\newblock
\shownote{\href{https://www.federalregister.gov/documents/2024/04/09/2024-07512/cybersecurity-in-the-marine-transportation-system}{federalregister.gov}}.


\bibitem[Organization({[n.\,d.]})]%
        {imo_regulations}
\bibfield{author}{\bibinfo{person}{International~Maritime Organization}.}
  \bibinfo{year}{[n.\,d.]}\natexlab{}.
\newblock \bibinfo{title}{Safety regulations for different types of ships}.
\newblock
\newblock
\newblock
\shownote{\href{https://www.imo.org/en/OurWork/Safety/Pages/RegulationsDefault.aspx}{imo.org}}.


\bibitem[Raunek(2023)]%
        {marine_insight_ship_types}
\bibfield{author}{\bibinfo{person}{Raunek}.} \bibinfo{year}{2023}\natexlab{}.
\newblock \bibinfo{title}{A Guide to Types of Ships}.
\newblock
\newblock
\newblock
\shownote{\href{https://www.marineinsight.com/guidelines/a-guide-to-types-of-ships/}{marineinsight.com}}.


\bibitem[Vance(2025)]%
        {britannica_ferries}
\bibfield{author}{\bibinfo{person}{James~E. Vance}.}
  \bibinfo{year}{2025}\natexlab{}.
\newblock \bibinfo{title}{Ferries}.
\newblock
\newblock
\newblock
\shownote{\href{https://www.britannica.com/technology/ship/Ferries}{britannica.com}}.


\bibitem[Center({[n.\,d.]})]%
        {houston_maritime_ship_types}
\bibfield{author}{\bibinfo{person}{Houston Maritime~Education Center}.}
  \bibinfo{year}{[n.\,d.]}\natexlab{}.
\newblock \bibinfo{title}{Types of Ships}.
\newblock
\newblock
\newblock
\shownote{\href{https://houstonmaritime.org/type-of-ships/}{houstonmaritime.org}}.


\bibitem[Register(2024)]%
        {lloyds_register_classification}
\bibfield{author}{\bibinfo{person}{Lloyd's Register}.}
  \bibinfo{year}{2024}\natexlab{}.
\newblock \bibinfo{title}{LR-RU-001 Rules and Regulations for the
  Classification of Ships}.
\newblock
\newblock
\newblock
\shownote{\href{https://www.lr.org/en/knowledge/lloyds-register-rules/rules-and-regulations-for-the-classification-of-ships/}{lr.org}}.


\bibitem[Group(2021)]%
        {oneocean_vessel_types}
\bibfield{author}{\bibinfo{person}{OneOcean Group}.}
  \bibinfo{year}{2021}\natexlab{}.
\newblock \bibinfo{title}{Vessel types explained}.
\newblock
\newblock
\newblock
\shownote{\href{https://www.oneocean.com/insights/vessel-types-explained}{oneocean.com}}.


\bibitem[Kerwin(2021)]%
        {suncam_ship_types}
\bibfield{author}{\bibinfo{person}{Kevin~M. Kerwin}.}
  \bibinfo{year}{2021}\natexlab{}.
\newblock \bibinfo{title}{Basic Ship Types \& Their Uses (Part 1)}.
\newblock
\newblock
\newblock
\shownote{\href{https://s3.amazonaws.com/suncam/docs/447.pdf\#page=2.00}{s3.amazonaws.com}}.


\bibitem[Group(2024)]%
        {mr_marine_ship_types}
\bibfield{author}{\bibinfo{person}{Mr.~Marine Group}.}
  \bibinfo{year}{2024}\natexlab{}.
\newblock \bibinfo{title}{Types of Ships - What are the differences?}
\newblock
\newblock
\newblock
\shownote{\href{https://mr-marinegroup.com/types-of-ships/}{mr-marinegroup.com}}.


\bibitem[of~Ships({[n.\,d.]})]%
        {history_of_ships}
\bibfield{author}{\bibinfo{person}{History of Ships}.}
  \bibinfo{year}{[n.\,d.]}\natexlab{}.
\newblock \bibinfo{title}{Different Types of Ships}.
\newblock
\newblock
\newblock
\shownote{\href{https://www.historyofships.net/ship-facts/types-of-ships/}{historyofships.net}}.


\bibitem[CareerExplorer({[n.\,d.]})]%
        {career_explorer}
\bibfield{author}{\bibinfo{person}{CareerExplorer}.}
  \bibinfo{year}{[n.\,d.]}\natexlab{}.
\newblock \bibinfo{title}{What does a merchant mariner do?}
\newblock
\newblock
\newblock
\shownote{\href{https://www.careerexplorer.com/careers/merchant-mariner/}{careerexplorer.com}}.


\bibitem[(MSC)({[n.\,d.]})]%
        {military_sealift_command}
\bibfield{author}{\bibinfo{person}{Military Sealift~Command (MSC)}.}
  \bibinfo{year}{[n.\,d.]}\natexlab{}.
\newblock \bibinfo{title}{Entry Level}.
\newblock
\newblock
\newblock
\shownote{\href{https://sealiftcommand.com/departments/entry-level}{sealiftcommand.com}}.


\bibitem[of~Technology and (MITAGS)(2024)]%
        {mitags}
\bibfield{author}{\bibinfo{person}{Maritime~Institute of Technology} {and}
  \bibinfo{person}{Graduate~Studies (MITAGS)}.}
  \bibinfo{year}{2024}\natexlab{}.
\newblock \bibinfo{title}{Maritime Jobs 101}.
\newblock
\newblock
\newblock
\shownote{\href{https://www.mitags.org/become-merchant-marine/}{mitags.org}}.


\bibitem[Training({[n.\,d.]})]%
        {maritime_professinal_training}
\bibfield{author}{\bibinfo{person}{Maritime~Professional Training}.}
  \bibinfo{year}{[n.\,d.]}\natexlab{}.
\newblock \bibinfo{title}{Merchant Careers}.
\newblock
\newblock
\newblock
\shownote{\href{https://www.mptusa.com/merchant-careers\#FAQ2}{mptusa.com}}.


\bibitem[Administration(2023)]%
        {dot_maritime_administration}
\bibfield{author}{\bibinfo{person}{U.S. DoT~Maritime Administration}.}
  \bibinfo{year}{2023}\natexlab{}.
\newblock \bibinfo{title}{Military to Mariner}.
\newblock
\newblock
\newblock
\shownote{\href{https://www.maritime.dot.gov/outreach/military-mariner\#My\%20title}{maritime.dot.gov}}.


\bibitem[Academy({[n.\,d.]})]%
        {northwest_maritime_academy}
\bibfield{author}{\bibinfo{person}{Northwest~Maritime Academy}.}
  \bibinfo{year}{[n.\,d.]}\natexlab{}.
\newblock \bibinfo{title}{The Path to Choosing Your Maritime Job}.
\newblock
\newblock
\newblock
\shownote{\href{https://northwestmaritimeacademy.com/path-choosing-maritime-job/}{northwestmaritimeacademy.com}}.


\bibitem[(NSCC)({[n.\,d.]})]%
        {nscc}
\bibfield{author}{\bibinfo{person}{Nova Scotia Community~College (NSCC)}.}
  \bibinfo{year}{[n.\,d.]}\natexlab{}.
\newblock \bibinfo{title}{Careers at Sea}.
\newblock
\newblock
\newblock
\shownote{\href{https://www.nscc.ca/careersatsea/career-options/ships-crew.asp}{nscc.ca}}.


\bibitem[Abdullah(2023)]%
        {marine_agency_services}
\bibfield{author}{\bibinfo{person}{Ashraful~Islam Abdullah}.}
  \bibinfo{year}{2023}\natexlab{}.
\newblock \bibinfo{title}{Crew Positions on a Ship}.
\newblock
\newblock
\newblock
\shownote{\href{https://www.mascrew.com/post/crew-positions-on-a-ship}{mascrew.com}}.


\bibitem[ShipFinex(2024)]%
        {shipfinex}
\bibfield{author}{\bibinfo{person}{ShipFinex}.}
  \bibinfo{year}{2024}\natexlab{}.
\newblock \bibinfo{title}{Ship Crew's Ranks, Positions \& Responsibilities}.
\newblock
\newblock
\newblock
\shownote{\href{https://www.shipfinex.com/blog/ship-crew-s-ranks-positions-responsibilities}{shipfinex.com}}.


\bibitem[Nautic(2024)]%
        {primonautic}
\bibfield{author}{\bibinfo{person}{Primo Nautic}.}
  \bibinfo{year}{2024}\natexlab{}.
\newblock \bibinfo{title}{Maritime Personnel \& Roles}.
\newblock
\newblock
\newblock
\shownote{\href{https://primonautic.com/blog/understanding-ship-hierarchies-ranks-and-roles-explained/}{primonautic.com}}.


\bibitem[Anchoredemia(2023)]%
        {anchoredemia}
\bibfield{author}{\bibinfo{person}{Anchoredemia}.}
  \bibinfo{year}{2023}\natexlab{}.
\newblock \bibinfo{title}{What positions usually make up the crew of a ship?}
\newblock
\newblock
\newblock
\shownote{\href{https://anclademia.com/en/blog/boat-crew/}{anclademia.com}}.


\bibitem[Ships({[n.\,d.]})]%
        {cslships}
\bibfield{author}{\bibinfo{person}{CSL Ships}.}
  \bibinfo{year}{[n.\,d.]}\natexlab{}.
\newblock \bibinfo{title}{Seafaring Roles and Responsibilities}.
\newblock
\newblock
\newblock
\shownote{\href{https://cslships.com/seafaring-roles-and-responsibilities/}{cslships.com}}.


\bibitem[Wade({[n.\,d.]})]%
        {life_of_sailing}
\bibfield{author}{\bibinfo{person}{Daniel Wade}.}
  \bibinfo{year}{[n.\,d.]}\natexlab{}.
\newblock \bibinfo{title}{Types of Sailors}.
\newblock
\newblock
\newblock
\shownote{\href{https://www.lifeofsailing.com/post/types-of-sailors}{lifeofsailing.com}}.


\bibitem[Bacaintan(2016)]%
        {ship_crew_structure}
\bibfield{author}{\bibinfo{person}{Narcis Bacaintan}.}
  \bibinfo{year}{2016}\natexlab{}.
\newblock \bibinfo{title}{Find your place in a ship crew structure}.
\newblock
\newblock
\newblock
\shownote{\href{https://www.linkedin.com/pulse/find-your-place-ship-crew-structure-narcis-bacaintan/}{linkedin.com}}.


\bibitem[Institution({[n.\,d.]})]%
        {woods_hole}
\bibfield{author}{\bibinfo{person}{Woods Hole~Oceanographic Institution}.}
  \bibinfo{year}{[n.\,d.]}\natexlab{}.
\newblock \bibinfo{title}{Ship Positions}.
\newblock
\newblock
\newblock
\shownote{\href{https://www.whoi.edu/what-we-do/explore/cruise-planning/cruise-planning-shipboard-at-sea/cruise-planning-ship-positions/}{whoi.edu}}.


\bibitem[Bacaintan(2016)]%
        {deck_department}
\bibfield{author}{\bibinfo{person}{Narcis Bacaintan}.}
  \bibinfo{year}{2016}\natexlab{}.
\newblock \bibinfo{title}{Crew structure on board merchant vessels - deck
  department}.
\newblock
\newblock
\newblock
\shownote{\href{https://www.linkedin.com/pulse/crew-structure-board-merchant-vessels-deck-department-bacaintan/}{linkedin.com}}.


\bibitem[Organization({[n.\,d.]})]%
        {imo}
\bibfield{author}{\bibinfo{person}{International~Maritime Organization}.}
  \bibinfo{year}{[n.\,d.]}\natexlab{}.
\newblock \bibinfo{title}{Introduction to IMO}.
\newblock
\newblock
\newblock
\shownote{\href{https://www.imo.org/en/About/Pages/Default.aspx}{imo.org}}.


\bibitem[Council({[n.\,d.]})]%
        {world_shipping_council}
\bibfield{author}{\bibinfo{person}{World~Shipping Council}.}
  \bibinfo{year}{[n.\,d.]}\natexlab{}.
\newblock \bibinfo{title}{Shipping Regulation}.
\newblock
\newblock
\newblock
\shownote{\href{https://www.worldshipping.org/shipping-regulation}{worldshipping.org}}.


\bibitem[ClearSeas(2024)]%
        {clearseas}
\bibfield{author}{\bibinfo{person}{ClearSeas}.}
  \bibinfo{year}{2024}\natexlab{}.
\newblock \bibinfo{title}{How is the Marine Shipping Industry Regulated?}
\newblock
\newblock
\newblock
\shownote{\href{https://clearseas.org/insights/marine-shipping-industry-regulated/}{clearseas.org}}.


\bibitem[DOT({[n.\,d.]})]%
        {maritime_and_waterways}
\bibfield{author}{\bibinfo{person}{U.S. DOT}.}
  \bibinfo{year}{[n.\,d.]}\natexlab{}.
\newblock \bibinfo{title}{Maritime and Waterways}.
\newblock
\newblock
\newblock
\shownote{\href{https://www.transportation.gov/maritime-and-waterways}{transportation.gov}}.


\bibitem[Ahmed(2022)]%
        {top_10_classification}
\bibfield{author}{\bibinfo{person}{Zahra Ahmed}.}
  \bibinfo{year}{2022}\natexlab{}.
\newblock \bibinfo{title}{Top 10 Classif. Societies In The World}.
\newblock
\newblock
\newblock
\shownote{\href{https://www.marineinsight.com/maritime-law/classification-societies-in-the-world/}{marineinsight.com}}.


\bibitem[Shipping(2023)]%
        {heisenbergshipping}
\bibfield{author}{\bibinfo{person}{Heisenberg Shipping}.}
  \bibinfo{year}{2023}\natexlab{}.
\newblock \bibinfo{title}{What Is Classification Society in Shipping?}
\newblock
\newblock
\newblock
\shownote{\href{https://heisenbergshipping.com/classification-society/}{heisenbergshipping.com}}.


\bibitem[Code(2025)]%
        {us_code}
\bibfield{author}{\bibinfo{person}{U.S. Code}.}
  \bibinfo{year}{2025}\natexlab{}.
\newblock \bibinfo{title}{§3316. Classification societies}.
\newblock
\newblock
\newblock
\shownote{\href{https://uscode.house.gov/view.xhtml?req=granuleid:USC-2000-title46-section3316\&num=0\&edition=2000}{uscode.house.gov}}.


\bibitem[of~Classification Societies~(IACS)({[n.\,d.]})]%
        {iacs}
\bibfield{author}{\bibinfo{person}{International~Association of Classification
  Societies~(IACS)}.} \bibinfo{year}{[n.\,d.]}\natexlab{}.
\newblock
\newblock
\newblock
\shownote{\href{https://iacs.org.uk/}{iacs.org.uk}}.


\bibitem[Guard({[n.\,d.]})]%
        {uscg_classification}
\bibfield{author}{\bibinfo{person}{U.S.~Coast Guard}.}
  \bibinfo{year}{[n.\,d.]}\natexlab{}.
\newblock \bibinfo{title}{Classification Society Authorizations}.
\newblock
\newblock
\newblock
\shownote{\href{https://www.dco.uscg.mil/Our-Organization/Assistant-Commandant-for-Prevention-Policy-CG-5P/Inspections-Compliance-CG-5PC-/Commercial-Vessel-Compliance/Flag-State-Control-Division/ClassSocAuth/}{dco.uscg.mil}}.


\bibitem[Ferns({[n.\,d.]})]%
        {ferns}
\bibfield{author}{\bibinfo{person}{B. Ferns}.}
  \bibinfo{year}{[n.\,d.]}\natexlab{}.
\newblock \bibinfo{title}{Class \& Flag Survey: Do you know the difference?}
\newblock
\newblock
\newblock
\shownote{\href{https://www.seaoc.co.uk/articles/class-amp-flag-survey-do-you-know-the-difference}{seaoc.co.uk}}.


\bibitem[Chopra(2021)]%
        {chopra}
\bibfield{author}{\bibinfo{person}{Karan Chopra}.}
  \bibinfo{year}{2021}\natexlab{}.
\newblock \bibinfo{title}{What are Flag States in the Shipping Industry And
  What’s Their Role?}
\newblock
\newblock
\newblock
\shownote{\href{https://www.marineinsight.com/maritime-law/what-are-flag-states-in-the-shipping-industry-2/}{marineinsight.com}}.


\bibitem[Flagge({[n.\,d.]})]%
        {deutsche_flagge}
\bibfield{author}{\bibinfo{person}{Deutsche Flagge}.}
  \bibinfo{year}{[n.\,d.]}\natexlab{}.
\newblock \bibinfo{title}{Flag States, Classification Societies}.
\newblock
\newblock
\newblock
\shownote{\href{https://www.deutsche-flagge.de/en/pscen/findings/ranking}{deutsche-flagge.de}}.


\bibitem[EMISA({[n.\,d.]})]%
        {emisa}
\bibfield{author}{\bibinfo{person}{EMISA}.}
  \bibinfo{year}{[n.\,d.]}\natexlab{}.
\newblock \bibinfo{title}{The role of Flag States}.
\newblock
\newblock
\newblock
\shownote{\href{https://emisa.eu/the-role-of-flag-states/}{emisa.eu}}.


\bibitem[of~Dredging Companies~(IADC)({[n.\,d.]})]%
        {iadc}
\bibfield{author}{\bibinfo{person}{International~Association of Dredging
  Companies~(IADC)}.} \bibinfo{year}{[n.\,d.]}\natexlab{}.
\newblock \bibinfo{title}{Regulatory Bodies, Agencies, Commissions and
  Organisations}.
\newblock
\newblock
\newblock
\shownote{\href{https://www.iadc-dredging.com/subject/regulatory-bodies-agencies-commissions/}{iadc-dredging.com}}.


\bibitem[Organization({[n.\,d.]})]%
        {solas}
\bibfield{author}{\bibinfo{person}{International~Maritime Organization}.}
  \bibinfo{year}{[n.\,d.]}\natexlab{}.
\newblock \bibinfo{title}{International Convention for the Safety of Life at
  Sea (SOLAS), 1974}.
\newblock
\newblock
\newblock
\shownote{\href{https://www.imo.org/en/About/Conventions/Pages/International-Convention-for-the-Safety-of-Life-at-Sea-(SOLAS),-1974.aspx}{imo.org}}.


\bibitem[Proceedings({[n.\,d.]})]%
        {isps}
\bibfield{author}{\bibinfo{person}{Coast~Guard Proceedings}.}
  \bibinfo{year}{[n.\,d.]}\natexlab{}.
\newblock \bibinfo{title}{ISPS/MTSA}.
\newblock
\newblock
\newblock
\shownote{\href{https://www.dco.uscg.mil/ISPS-MTSA/}{dco.uscg.mil}}.


\bibitem[Organization(2021)]%
        {maritime_cyber_risk}
\bibfield{author}{\bibinfo{person}{International~Maritime Organization}.}
  \bibinfo{year}{2021}\natexlab{}.
\newblock \bibinfo{title}{Maritime cyber risk}.
\newblock
\newblock
\newblock
\shownote{\href{https://www.imo.org/en/OurWork/Security/Pages/Cyber-security.aspx}{imo.org}}.


\bibitem[{International Maritime Organization}(2022)]%
        {imo_guidelines}
\bibfield{author}{\bibinfo{person}{{International Maritime Organization}}.}
  \bibinfo{year}{2022}\natexlab{}.
\newblock \bibinfo{title}{MSC-FAL.1/Circ.3/Rev.2 – Guidelines on Maritime
  Cyber Risk Management}.
\newblock
\newblock
\newblock
\shownote{\href{https://wwwcdn.imo.org/localresources/en/OurWork/Security/Documents/MSC-FAL.1-Circ.3-Rev.2\%20-\%20Guidelines\%20On\%20Maritime\%20Cyber\%20Risk\%20Management\%20(Secretariat)\%20(1).pdf}{wwwcdn.imo.org}}.


\bibitem[BIMCO et~al\mbox{.}(2024)]%
        {cyber_guidelines_v5}
\bibfield{author}{\bibinfo{person}{BIMCO}, \bibinfo{person}{CLIA},
  \bibinfo{person}{ICS}, \bibinfo{person}{INTERCARGO},
  \bibinfo{person}{INTERTANKO}, \bibinfo{person}{OCIMF}, {and}
  \bibinfo{person}{IUMI}.} \bibinfo{year}{2024}\natexlab{}.
\newblock \bibinfo{title}{The Guidelines on Cyber Security Onboard Ships}.
\newblock
\newblock
\newblock
\shownote{\href{https://www.maritimeglobalsecurity.org/media/g3qlxdaw/2024-11-14-guidelines_on_cyber_security-v5-final.pdf}{maritimeglobalsecurity.org}}.


\bibitem[Stouffer et~al\mbox{.}(2023)]%
        {nist80082r3}
\bibfield{author}{\bibinfo{person}{Keith Stouffer}, \bibinfo{person}{Victoria
  Pillitteri}, \bibinfo{person}{Susan Lightman}, \bibinfo{person}{Marshall
  Abrams}, {and} \bibinfo{person}{Adam Hahn}.} \bibinfo{year}{2023}\natexlab{}.
\newblock \bibinfo{title}{Guide to Operational Technology (OT) Security}.
\newblock
\newblock


\bibitem[Sahin et~al\mbox{.}(2023)]%
        {sahin2023investigating}
\bibfield{author}{\bibinfo{person}{Sena Sahin}, \bibinfo{person}{Suood
  Al~Roomi}, \bibinfo{person}{Tara Poteat}, {and} \bibinfo{person}{Frank Li}.}
  \bibinfo{year}{2023}\natexlab{}.
\newblock \showarticletitle{Investigating the Password Policy Practices of
  Website Administrators}. In \bibinfo{booktitle}{\emph{IEEE S\&P}}.
\newblock


\bibitem[Yong~Wong et~al\mbox{.}(2021)]%
        {yong2021inside}
\bibfield{author}{\bibinfo{person}{Miuyin Yong~Wong}, \bibinfo{person}{Matthew
  Landen}, \bibinfo{person}{Manos Antonakakis}, \bibinfo{person}{Douglas~M
  Blough}, \bibinfo{person}{Elissa~M Redmiles}, {and} \bibinfo{person}{Mustaque
  Ahamad}.} \bibinfo{year}{2021}\natexlab{}.
\newblock \showarticletitle{An inside look into the practice of malware
  analysis}. In \bibinfo{booktitle}{\emph{CCS}}.
\newblock


\bibitem[Wong et~al\mbox{.}(2024)]%
        {wong2024comparing}
\bibfield{author}{\bibinfo{person}{Miuyin~Yong Wong}, \bibinfo{person}{Matthew
  Landen}, \bibinfo{person}{Frank Li}, \bibinfo{person}{Fabian Monrose}, {and}
  \bibinfo{person}{Mustaque Ahamad}.} \bibinfo{year}{2024}\natexlab{}.
\newblock \showarticletitle{Comparing Malware Evasion Theory with Practice:
  Results from Interviews with Expert Analysts}. In
  \bibinfo{booktitle}{\emph{SOUPS}}.
\newblock


\bibitem[Akgul et~al\mbox{.}(2023)]%
        {akgul2023bug}
\bibfield{author}{\bibinfo{person}{Omer Akgul}, \bibinfo{person}{Taha
  Eghtesad}, \bibinfo{person}{Amit Elazari}, \bibinfo{person}{Omprakash
  Gnawali}, \bibinfo{person}{Jens Grossklags}, \bibinfo{person}{Michelle~L
  Mazurek}, \bibinfo{person}{Daniel Votipka}, {and} \bibinfo{person}{Aron
  Laszka}.} \bibinfo{year}{2023}\natexlab{}.
\newblock \showarticletitle{Bug Hunters’ Perspectives on the Challenges and
  Benefits of the Bug Bounty Ecosystem}. In \bibinfo{booktitle}{\emph{USENIX
  Security}}.
\newblock


\bibitem[Votipka et~al\mbox{.}(2018)]%
        {votipka2018hackers}
\bibfield{author}{\bibinfo{person}{Daniel Votipka}, \bibinfo{person}{Rock
  Stevens}, \bibinfo{person}{Elissa Redmiles}, \bibinfo{person}{Jeremy Hu},
  {and} \bibinfo{person}{Michelle Mazurek}.} \bibinfo{year}{2018}\natexlab{}.
\newblock \showarticletitle{Hackers vs. testers: A comparison of software
  vulnerability discovery processes}. In \bibinfo{booktitle}{\emph{IEEE S\&P}}.
\newblock


\bibitem[Gallardo et~al\mbox{.}(2024)]%
        {gallardo2024interdisciplinary}
\bibfield{author}{\bibinfo{person}{Andrea Gallardo}, \bibinfo{person}{Robert
  Erbes}, \bibinfo{person}{Katya Le~Blanc}, \bibinfo{person}{Lujo Bauer}, {and}
  \bibinfo{person}{Lorrie~Faith Cranor}.} \bibinfo{year}{2024}\natexlab{}.
\newblock \showarticletitle{Interdisciplinary Approaches to Cybervulnerability
  Impact Assessment for Energy Critical Infrastructure}. In
  \bibinfo{booktitle}{\emph{CHI}}.
\newblock


\bibitem[Evripidou et~al\mbox{.}(2023)]%
        {evripidou2023exploring}
\bibfield{author}{\bibinfo{person}{Stefanos Evripidou},
  \bibinfo{person}{Uchenna~D. Ani}, \bibinfo{person}{Stephen Hailes}, {and}
  \bibinfo{person}{Jeremy D.~McK. Watson}.} \bibinfo{year}{2023}\natexlab{}.
\newblock \showarticletitle{Exploring the Security Culture of Operational
  Technology ({OT}) Organisations: The Role of External Consultancy in
  Overcoming Organisational Barriers}. In \bibinfo{booktitle}{\emph{SOUPS}}.
\newblock


\bibitem[Li et~al\mbox{.}(2024)]%
        {li2024windows98}
\bibfield{author}{\bibinfo{person}{Karen Li}, \bibinfo{person}{Kopo~M.
  Ramokapane}, {and} \bibinfo{person}{Awais Rashid}.}
  \bibinfo{year}{2024}\natexlab{}.
\newblock \showarticletitle{``Yeah, it does have a\ldots{} Windows '98 Vibe'':
  Usability Study of Security Features in Programmable Logic Controllers}. In
  \bibinfo{booktitle}{\emph{EuroUSEC}}.
\newblock


\bibitem[Nunes et~al\mbox{.}(2024)]%
        {nunes2024exploiting}
\bibfield{author}{\bibinfo{person}{Matthew Nunes}, \bibinfo{person}{Hakan
  Kayan}, \bibinfo{person}{Pete Burnap}, \bibinfo{person}{Charith Perera},
  {and} \bibinfo{person}{Jason Dykes}.} \bibinfo{year}{2024}\natexlab{}.
\newblock \showarticletitle{Exploiting user-centred design to secure industrial
  control systems}. In \bibinfo{booktitle}{\emph{Frontiers in the Internet of
  Things}}.
\newblock


\bibitem[Fl{\r{a}} et~al\mbox{.}(2024)]%
        {fla2024challenges}
\bibfield{author}{\bibinfo{person}{Lars~Halvdan Fl{\r{a}}},
  \bibinfo{person}{Christoph~Alexander Thieme}, \bibinfo{person}{Martin~Gilje
  Jaatun}, {and} \bibinfo{person}{Geir~Kjetil Hanssen}.}
  \bibinfo{year}{2024}\natexlab{}.
\newblock \showarticletitle{Cybersecurity Challenges in Industrial Control
  Systems: An Interview Study with Asset Owners in Norway}. In
  \bibinfo{booktitle}{\emph{ESORICS}}.
\newblock


\bibitem[Evripidou and Watson(2022)]%
        {evripidou2022understanding}
\bibfield{author}{\bibinfo{person}{Stefanos Evripidou} {and}
  \bibinfo{person}{Jeremy D.~McK. Watson}.} \bibinfo{year}{2022}\natexlab{}.
\newblock \showarticletitle{Understanding Operational Technology Personnel’s
  Mindsets and Their Effect on Cybersecurity Perceptions: A Qualitative Study
  With Operational Technology Cybersecurity Practitioners}. In
  \bibinfo{booktitle}{\emph{USEC}}.
\newblock


\bibitem[Progoulakis et~al\mbox{.}(2021)]%
        {progoulakis2021cyber}
\bibfield{author}{\bibinfo{person}{Iosif Progoulakis}, \bibinfo{person}{Paul
  Rohmeyer}, {and} \bibinfo{person}{Nikita Nikitakos}.}
  \bibinfo{year}{2021}\natexlab{}.
\newblock \showarticletitle{Cyber physical systems security for maritime
  assets}. In \bibinfo{booktitle}{\emph{Journal of Marine Science and Eng.}}
\newblock


\bibitem[DiRenzo et~al\mbox{.}(2015)]%
        {direnzo2015little}
\bibfield{author}{\bibinfo{person}{Joseph DiRenzo}, \bibinfo{person}{Dana~A
  Goward}, {and} \bibinfo{person}{Fred~S Roberts}.}
  \bibinfo{year}{2015}\natexlab{}.
\newblock \showarticletitle{The little-known challenge of maritime cyber
  security}. In \bibinfo{booktitle}{\emph{IISA}}.
\newblock


\bibitem[Tran et~al\mbox{.}(2021)]%
        {tran2021marine}
\bibfield{author}{\bibinfo{person}{Ky Tran}, \bibinfo{person}{Sid Keene},
  \bibinfo{person}{Erik Fretheim}, {and} \bibinfo{person}{Michail
  Tsikerdekis}.} \bibinfo{year}{2021}\natexlab{}.
\newblock \showarticletitle{Marine network protocols and security risks}. In
  \bibinfo{booktitle}{\emph{Journal of Cybersecurity and Privacy}}.
\newblock


\bibitem[Amro et~al\mbox{.}(2022)]%
        {amro2022navigation}
\bibfield{author}{\bibinfo{person}{Ahmed Amro}, \bibinfo{person}{Aybars Oruc},
  \bibinfo{person}{Vasileios Gkioulos}, {and} \bibinfo{person}{Sokratis
  Katsikas}.} \bibinfo{year}{2022}\natexlab{}.
\newblock \showarticletitle{Navigation data anomaly analysis and detection}. In
  \bibinfo{booktitle}{\emph{Information}}.
\newblock


\bibitem[Lampe(2024)]%
        {lampe2024application}
\bibfield{author}{\bibinfo{person}{Benjamin Lampe}.}
  \bibinfo{year}{2024}\natexlab{}.
\newblock \showarticletitle{On the Application of Cyber-Informed Engineering
  (CIE)}. In \bibinfo{booktitle}{\emph{TPS-ISA}}.
\newblock


\bibitem[Organization(2021)]%
        {women_in_maritime}
\bibfield{author}{\bibinfo{person}{International~Maritime Organization}.}
  \bibinfo{year}{2021}\natexlab{}.
\newblock \bibinfo{title}{Women in Maritime}.
\newblock
\newblock
\newblock
\shownote{\href{https://www.imo.org/en/ourwork/technicalcooperation/pages/womeninmaritime.aspx}{imo.org}}.


\bibitem[Srivastava and Hopwood(2009)]%
        {srivastava2009practical}
\bibfield{author}{\bibinfo{person}{Prachi Srivastava} {and}
  \bibinfo{person}{Nick Hopwood}.} \bibinfo{year}{2009}\natexlab{}.
\newblock \showarticletitle{A practical iterative framework for qualitative
  data analysis}.
\newblock \bibinfo{journal}{\emph{International journal of qualitative
  methods}} (\bibinfo{year}{2009}).
\newblock


\bibitem[Raymaker(2025)]%
        {osf_codebook}
\bibfield{author}{\bibinfo{person}{Anna Raymaker}.}
  \bibinfo{year}{2025}\natexlab{}.
\newblock \bibinfo{title}{Codebook - OSF Repository}.
\newblock
\newblock
\newblock
\shownote{\href{https://osf.io/uyxg2/files/osfstorage?view_only=c4d12edeaac24c0189d655042866740d}{osf.io}}.


\bibitem[Fleiss et~al\mbox{.}(2013)]%
        {fleiss2013statistical}
\bibfield{author}{\bibinfo{person}{Joseph~L Fleiss}, \bibinfo{person}{Bruce
  Levin}, {and} \bibinfo{person}{Myunghee~Cho Paik}.}
  \bibinfo{year}{2013}\natexlab{}.
\newblock \bibinfo{booktitle}{\emph{Statistical methods for rates and
  proportions}}.
\newblock \bibinfo{publisher}{John Wiley \& Sons}.
\newblock


\bibitem[Guest et~al\mbox{.}(2020)]%
        {guest2020simple}
\bibfield{author}{\bibinfo{person}{Greg Guest}, \bibinfo{person}{Emily Namey},
  {and} \bibinfo{person}{Mario Chen}.} \bibinfo{year}{2020}\natexlab{}.
\newblock \showarticletitle{A simple method to assess and report thematic
  saturation in qualitative research}.
\newblock \bibinfo{journal}{\emph{PloS one}} (\bibinfo{year}{2020}).
\newblock


\bibitem[Bartsch(2011)]%
        {6046004}
\bibfield{author}{\bibinfo{person}{Steffen Bartsch}.}
  \bibinfo{year}{2011}\natexlab{}.
\newblock \showarticletitle{Practitioners' Perspectives on Security in Agile
  Development}. In \bibinfo{booktitle}{\emph{ARES}}.
\newblock


\bibitem[Haney et~al\mbox{.}(2018)]%
        {219400}
\bibfield{author}{\bibinfo{person}{Julie~M. Haney}, \bibinfo{person}{Mary
  Theofanos}, \bibinfo{person}{Yasemin Acar}, {and}
  \bibinfo{person}{Sandra~Spickard Prettyman}.}
  \bibinfo{year}{2018}\natexlab{}.
\newblock \showarticletitle{"We make it a big deal in the company": Security
  Mindsets in Organizations that Develop Cryptographic Products}. In
  \bibinfo{booktitle}{\emph{SOUPS}}.
\newblock


\bibitem[Acar et~al\mbox{.}(2017a)]%
        {7958576}
\bibfield{author}{\bibinfo{person}{Yasemin Acar}, \bibinfo{person}{Michael
  Backes}, \bibinfo{person}{Sascha Fahl}, \bibinfo{person}{Simson Garfinkel},
  \bibinfo{person}{Doowon Kim}, \bibinfo{person}{Michelle~L. Mazurek}, {and}
  \bibinfo{person}{Christian Stransky}.} \bibinfo{year}{2017}\natexlab{a}.
\newblock \showarticletitle{Comparing the Usability of Cryptographic APIs}. In
  \bibinfo{booktitle}{\emph{IEEE S\&P}}.
\newblock


\bibitem[Acar et~al\mbox{.}(2017b)]%
        {205176}
\bibfield{author}{\bibinfo{person}{Yasemin Acar}, \bibinfo{person}{Christian
  Stransky}, \bibinfo{person}{Dominik Wermke}, \bibinfo{person}{Michelle~L.
  Mazurek}, {and} \bibinfo{person}{Sascha Fahl}.}
  \bibinfo{year}{2017}\natexlab{b}.
\newblock \showarticletitle{Security Developer Studies with {GitHub} Users:
  Exploring a Convenience Sample}. In \bibinfo{booktitle}{\emph{SOUPS}}.
\newblock


\bibitem[Derr et~al\mbox{.}(2017)]%
        {derr2017keep}
\bibfield{author}{\bibinfo{person}{Erik Derr}, \bibinfo{person}{Sven Bugiel},
  \bibinfo{person}{Sascha Fahl}, \bibinfo{person}{Yasemin Acar}, {and}
  \bibinfo{person}{Michael Backes}.} \bibinfo{year}{2017}\natexlab{}.
\newblock \showarticletitle{Keep me updated: An empirical study of third-party
  library updatability on android}. In \bibinfo{booktitle}{\emph{CCS}}.
\newblock


\bibitem[Voronkov et~al\mbox{.}(2019)]%
        {voronkov2019system}
\bibfield{author}{\bibinfo{person}{Artem Voronkov}, \bibinfo{person}{Leonardo~A
  Martucci}, {and} \bibinfo{person}{Stefan Lindskog}.}
  \bibinfo{year}{2019}\natexlab{}.
\newblock \showarticletitle{System administrators prefer command line
  interfaces, don't they? an exploratory study of firewall interfaces}. In
  \bibinfo{booktitle}{\emph{SOUPS}}.
\newblock


\bibitem[Haney and Lutters(2018)]%
        {haney2018s}
\bibfield{author}{\bibinfo{person}{Julie~M Haney} {and}
  \bibinfo{person}{Wayne~G Lutters}.} \bibinfo{year}{2018}\natexlab{}.
\newblock \showarticletitle{"It's Scary...It's...Confusing...It's Dull": How
  Cybersecurity Advocates Overcome Negative Perceptions of Security}. In
  \bibinfo{booktitle}{\emph{SOUPS}}.
\newblock


\bibitem[Gorski et~al\mbox{.}(2018)]%
        {gorski2018developers}
\bibfield{author}{\bibinfo{person}{Peter~Leo Gorski}, \bibinfo{person}{Luigi~Lo
  Iacono}, \bibinfo{person}{Dominik Wermke}, \bibinfo{person}{Christian
  Stransky}, \bibinfo{person}{Sebastian M{\"o}ller}, \bibinfo{person}{Yasemin
  Acar}, {and} \bibinfo{person}{Sascha Fahl}.} \bibinfo{year}{2018}\natexlab{}.
\newblock \showarticletitle{Developers deserve security warnings, too: On the
  effect of integrated security advice on cryptographic API misuse}. In
  \bibinfo{booktitle}{\emph{SOUPS}}.
\newblock


\bibitem[Gerlitz et~al\mbox{.}(2021)]%
        {gerlitz2021please}
\bibfield{author}{\bibinfo{person}{Eva Gerlitz}, \bibinfo{person}{Maximilian
  H{\"a}ring}, {and} \bibinfo{person}{Matthew Smith}.}
  \bibinfo{year}{2021}\natexlab{}.
\newblock \showarticletitle{Please do not use! \_ or your license plate number:
  Analyzing password policies in german companies}. In
  \bibinfo{booktitle}{\emph{SOUPS}}.
\newblock


\bibitem[Bruckman(2024)]%
        {bruckman2024compensation}
\bibfield{author}{\bibinfo{person}{Amy Bruckman}.}
  \bibinfo{year}{2024}\natexlab{}.
\newblock \bibinfo{title}{Should You Compensate Research Study Participants?}
\newblock
\newblock
\newblock
\shownote{\href{https://asbruckman.medium.com/should-you-compensate-research-study-participants-d6ad34c4babc}{asbruckman.medium.com}}.


\bibitem[Military.com(2024)]%
        {military2024navy}
\bibfield{author}{\bibinfo{person}{Military.com}.}
  \bibinfo{year}{2024}\natexlab{}.
\newblock \bibinfo{title}{Navy Chief Demoted After Installing Unauthorized
  Satellite Dish on Warship to Access Internet}.
\newblock
\newblock
\newblock
\shownote{\href{https://www.military.com/daily-news/2024/09/06/navy-officer-demoted-after-installing-unauthorized-satellite-dish-warship-access-internet.html}{military.com}}.


\bibitem[Ho et~al\mbox{.}(2025)]%
        {ho2025phishing}
\bibfield{author}{\bibinfo{person}{Grant Ho}, \bibinfo{person}{Ariana Mirian},
  \bibinfo{person}{Elisa Luo}, \bibinfo{person}{Khang Tong},
  \bibinfo{person}{Euyhyun Lee}, \bibinfo{person}{Lin Liu},
  \bibinfo{person}{Christopher~A. Longhurst}, \bibinfo{person}{Christian
  Dameff}, \bibinfo{person}{Stefan Savage}, {and} \bibinfo{person}{Geoffrey~M.
  Voelker}.} \bibinfo{year}{2025}\natexlab{}.
\newblock \showarticletitle{Understanding the Efficacy of Phishing Training in
  Practice}. In \bibinfo{booktitle}{\emph{IEEE S\&P}}.
\newblock


\bibitem[Sasse et~al\mbox{.}(2022)]%
        {sasse2022rebooting}
\bibfield{author}{\bibinfo{person}{M~Angela Sasse}, \bibinfo{person}{Jonas
  Hielscher}, \bibinfo{person}{Jennifer Friedauer}, {and}
  \bibinfo{person}{Annalina Buckmann}.} \bibinfo{year}{2022}\natexlab{}.
\newblock \showarticletitle{Rebooting IT security awareness--how organisations
  can encourage and sustain secure behaviours}. In
  \bibinfo{booktitle}{\emph{ESORICS}}.
\newblock


\bibitem[DHS({[n.\,d.]})]%
        {us_dhs_infrastructure}
\bibfield{author}{\bibinfo{person}{U.S. DHS}.}
  \bibinfo{year}{[n.\,d.]}\natexlab{}.
\newblock \bibinfo{title}{Critical Infrastructure Sectors}.
\newblock
\newblock
\newblock
\shownote{\href{https://www.cisa.gov/topics/critical-infrastructure-security-and-resilience/critical-infrastructure-sectors}{cisa.gov}}.


\bibitem[{North American Electric Reliability Corporation}(2024)]%
        {nerc_cip}
\bibfield{author}{\bibinfo{person}{{North American Electric Reliability
  Corporation}}.} \bibinfo{year}{2024}\natexlab{}.
\newblock \bibinfo{title}{NERC Critical Infrastructure Protection (CIP)
  Standards}.
\newblock
\newblock
\newblock
\shownote{\href{https://www.nerc.com/pa/Stand/Pages/ReliabilityStandards.aspx}{nerc.com}}.


\bibitem[Organization(2017)]%
        {imo_msc428}
\bibfield{author}{\bibinfo{person}{International~Maritime Organization}.}
  \bibinfo{year}{2017}\natexlab{}.
\newblock \bibinfo{title}{Resolution MSC.428(98) - Maritime Cyber Risk
  Management in Safety Management Systems}.
\newblock
\newblock
\newblock
\shownote{\href{https://wwwcdn.imo.org/localresources/en/OurWork/Security/Documents/Resolution\%20MSC.428(98).pdf}{wwwcdn.imo.org}}.


\end{thebibliography}

\appendix

\section{Appendix}
\label{sec:appendix}

\subsection{Survey Questions}
\label{sec:survey}

\begin{enumerate}
    \item What is (or was) your official role (job title) on the ship?
    \item Are you currently employed in this shipping position? If not, please write when you retired from this position.
    \item How long have you worked aboard ships (if less than 1 year, pu\new{t} answer in months)?
    \item Have you served in any other positions on ships (other than the role selected above)? If so, what and for how long?
    \item How large is (or was) the ship you serve on?
    \item What is (or was) the primary function of the ship?
    \item If your ship transports goods or cargo, please list the kind(s) of goods or cargo it transports.
    \item How many crew members (including yourself) serve(d) on your ship? If you don't know the exact number, please try to estimate a range or value.
    \item What organizational standards does your ship follow, specifically for safety and security? Select all that apply. For any that are not covered by the options available, please list them under the "Other" option.
    \item \new{**Asked by email after the interview to gather more demographic data** Please select your age from the following ranges or select "I do not wish to share this information":}
    \begin{enumerate}
        \new{\item Under 20 years old
        \item 20-29 years old
        \item 30-39 years old
        \item 40-49 years old
        \item 50-59 years old
        \item 60-69 years old
        \item 70-79 years old
        \item 80-89 years old
        \item 90 years or older
        \item I do not wish to share this information}
    \end{enumerate}
\end{enumerate}

\subsection{Interview Questions}
\label{sec:interview}
Below is our list of interview questions and the statements the interviewer gave prior to the start of each section of the interview. Each interview was conducted using the same questions and asked by the same person for consistency.

\subsubsection{Beginning Statement}

Before we begin with the questioning, I want to remind you that everything you say will be completely anonymous. I will record our voices during this conversation to remind myself later of specific details. That recording will be stored on an encrypted hard drive, anonymized, and destroyed after I perform an analysis of it. To confirm, is it ok if I begin the audio recording over Zoom?

In this study, we have a set number of questions that have to be asked to everyone in the same way and same order. At some point in the interview, you may feel that I am asking a question you have already answered because sometimes people naturally answer questions I haven not asked yet. I apologize for that in advance, but I still have to ask the question to confirm your answer and to ensure everyone receives the same interview. This helps us limit bias as researchers.

If you are confused about a question, feel free to ask for additional information. Do you have any questions before we begin?

\subsubsection{Background}
First, I would like to know a little more about you and your position.

\begin{enumerate}
    \item What are your main responsibilities in your role as X on the ship?
\end{enumerate}

\subsubsection{General Security Questions}
Now, I am going to ask general questions about security on the ship.
\begin{enumerate}
    \setcounter{enumi}{1}
    \item When you are working on a ship, what do you think are the most significant threats to ship security and operations?
    \item Are there any times during shipping trips that you feel more vulnerable to threats? Why do you feel this way?
    \item Who is responsible for the general security of the ship?
    \item Are there any measures you take as a crew to help protect against threats to the ship?
    \item Are there times during ship operations when people board your ship that are not part of your crew?
    \begin{enumerate}
        \item IF YES TO ABOVE: Are there any specific protection measures you take when third parties like this board the ship?
    \end{enumerate}
    \item Are there any specific protection measures you take when approaching a port? Are these measures different depending on the port you are approaching?
    \item Are there any specific protection measures you take when in the open ocean, far from civilization?\item Are there any specific protection measures you take when in a crowded waterway, near lots of other vessels?
\end{enumerate}

\subsubsection{Cybersecurity Practices and Incidents Questions}
Now, I am going to ask about devices and how you handle issues on ship.
\begin{enumerate}
    \setcounter{enumi}{9}
    \item What devices or equipment on the ship are you most afraid of malfunctioning or becoming unavailable?
    \item Of the devices you just mentioned, do you know what to do if they malfunction?
    \item If you had to take all your navigation devices offline during an emergency, do you feel confident that the crew has the necessary training and skills to continue to move the ship from point A to point B?
\end{enumerate}

Now, I am going to ask some cybersecurity and cyberattack-specific questions.

\begin{enumerate}
    \setcounter{enumi}{12}
    \item When you hear the phrase 'cybersecurity for ships', what comes to mind? 
    \item Have you experienced any cybersecurity issues or incidents aboard a marine vessel? If so, what happened and how was it handled?
    \item Do you feel confident in your ability to handle cyber-attacks?
    \begin{enumerate}
        \item IF CONFIDENT: Is there a specific response plan in place on board for dealing with cyber-attacks or events? If so, can you describe what that plan involves?
        \item IF NOT CONFIDENT: What makes you feel less confident in handling cyber-attacks?
    \end{enumerate}
    \item Are there any devices that you worry could be affected by a cyber-attack?
\end{enumerate}

The next questions involve cybersecurity training you may have received.

\begin{enumerate}
    \setcounter{enumi}{16}
    \item Have you ever received cybersecurity training for your job on the ship?
    \begin{enumerate}
        \item IF YES TO ABOVE: Can you describe the cybersecurity training you received for your role on the ship? What did it involve and what did you learn?
        \item IF YES TO ABOVE: Do you feel the cybersecurity training you received was useful in preparing you? Is there anything that you feel was missing or that you still feel unprepared for?
    \end{enumerate}
\end{enumerate}

\subsubsection{Comparative Cybersecurity Questions}
These next questions are very similar to the general security threat questions but focused on cybersecurity instead. 

\begin{enumerate}
    \setcounter{enumi}{17}
    \item When you are working on a ship, what do you think are the most significant cybersecurity threats?
    \item Are there any times during shipping trips that you feel more vulnerable to cybersecurity threats? Why do you feel this way?
    \item Who is responsible for the cybersecurity of the ship?
    \item Are there any measures you take as a crew to help protect against cybersecurity threats to the ship?
    \item Think about times when people other than your crew come aboard the ship. Are there any specific cybersecurity protection measures you take when third parties like this board the ship?
    \item Are there any specific cybersecurity protection measures you take when approaching a port? Are these measures different depending on the port you are approaching?
    \item Are there any specific cybersecurity protection measures you take when in the open ocean, far from civilization?
    \item Are there any specific cybersecurity protection measures you take when in a crowded waterway, near lots of other vessels?	
\end{enumerate}

\subsubsection{Regulation and Standards Questions}
If you recall, I had you select some of the organizations and standards you follow on the ship for safety and security, like SOLAS, ISM, and ISPS. Now, I am going to ask some questions about the standards and regulations you follow.

\begin{enumerate}
    \setcounter{enumi}{25}
    \item You mentioned following Flag State safety regulations. What nations have you followed standards for in your position?
    \item You mentioned using X, Y, and Z standards for safety and security. What do you believe are the main benefits of following these standards on your ship?
    \item Are there any challenges or drawbacks you've experienced with these safety and security standards
    \item In your opinion, how could these safety and security standards be improved to better support your work?
\end{enumerate}

Now, I am going to ask questions more specifically about cybersecurity standards for ships.

\begin{enumerate}
    \setcounter{enumi}{29}
    \item As part of many safety and security standards for ships, there are now also cybersecurity recommendations and requirements. Are you familiar with cybersecurity standards for ships?
    \begin{enumerate}
        \item IF YES TO ABOVE: What do you believe are the main benefits of following cybersecurity standards on ship?
        \item Are there any challenges or drawbacks you've experienced with cybersecurity standards?
        \item In your opinion, how could these cybersecurity standards be improved to better support your work?
    \end{enumerate}
    \begin{enumerate}
        \item IF NO TO ABOVE: Hypothetically, what do you believe are the main benefits of following cybersecurity standards on ship?
        \item Are there any challenges or drawbacks you've experienced when following any cybersecurity rules on ship?
        \item In your opinion, how could these cybersecurity rules be improved to better support your work?
    \end{enumerate}

\end{enumerate}

\subsubsection{Final Statement}
That is the end of our questions. Thank you so much for giving us your time. We will compile all participant responses and write a paper with our analysis. If you want, I can send you a copy of the paper when it is published. Additionally, we are still looking for more participants, so if you have any friends or colleagues that you think may fit our needs, we would love to include them. If you have them fill out the survey, that would be extremely helpful. I can resend the survey link to you if you have people to distribute it to.

\subsection{Extended Background Information}
\label{sec:ship_details}
\noindent\textbf{Ships and Their Classifications.} \emph{Cargo ships} transport goods, ranging from dry cargo (e.g., containers, grains) to liquid cargo (e.g., crude oil, LNG). \emph{Passenger ships}, such as ferries and cruise liners, serve travel and tourism needs. \emph{Special-purpose vessels}, including research ships, offshore supply vessels, cable ships, tugboats, and buoy tenders, support distinct maritime operations. A subset of these special-purpose vessels, \emph{Workboats}, such as tugboats and buoy tenders, play critical roles in supporting larger vessels and maintaining maritime infrastructure. By including these specialized vessels, our study highlights the diverse operations within the maritime sector. 

\noindent\textbf{Mariners and Their Roles.}
Mariners are responsible for ensuring ship safety, operational efficiency, and compliance with regulations. Key roles include \emph{Captains (Masters)}, who hold ultimate responsibility for the ship’s operations; \emph{Chief Mates}, who oversee deck operations; and \emph{Chief Engineers}, who manage propulsion and machinery. Supporting officers and crew contribute specialized expertise in navigation, maintenance, and safety. Crew composition varies by vessel type, with larger ships requiring more personnel to manage their increased complexity.

This study focuses on mariners responsible for ship security and cybersecurity, capturing insights from officer-level positions such as Captains, Chief Mates, and Engineers, as well as their military equivalents (e.g., Deck Watch Officer in the Coast Guard). Their oversight of critical systems highlights the importance of understanding cybersecurity challenges in maritime operations.

\subsection{Pilot Study}
\label{sec:pilot}

Our pilot study consisted of two rounds, with changes to the interview question set implemented after each round. Demographic information for the pilot participants, Pilot Participant 1 (PP1) and Pilot Participant 2 (PP2), is summarized in the first two rows of Table ~\ref{tab:participant_info}. Insights gained from the pilot study informed the refinement of our final interview design introduced in the previous subsection. \\
\textbf{Round 1:} In the first round, one participant was interviewed using a question set focused solely on cybersecurity and cyber-attack topics. While these questions aimed to provide insights into the participant’s understanding of cybersecurity threats, they proved too narrow in scope. Specifically, the participant, who had limited cybersecurity knowledge, interpreted “cybersecurity” as primarily involving “computers, malware, or phishing.” This restricted the depth of responses and overlooked other critical aspects of cybersecurity, such as physical access threats. For instance, during our background research with industry professionals, physical access was identified as a significant threat. This critical context would have been missed if interviewees framed cybersecurity solely in terms of technical vulnerabilities. The limitations of this approach led to significant revisions in the question set. Due to the substantial changes made and the limited usefulness of the responses, this interview was not coded or included in the final results. \\
\textbf{Round 2:} The second round involved one participant and included a revised question set that removed several cybersecurity threat questions. However, this omission led to insufficient coverage of cybersecurity threats, particularly in comparison to physical threats. As a result, additional “mirrored” cybersecurity questions were developed after this round to ensure meaningful responses could be gathered from participants with varying levels of expertise. Although this interview was similar enough to the final question set to be coded and included in the final results, it was not used to determine saturation. Instead, its purpose was to gather insights that informed the refinement of the final question set. Saturation calculations began with Participant 1 in the main study.

\subsection{\new{Additional Demographics Data}}
\label{sec:demo_data}
\begin{table}[H]
\centering
\small
\begin{tabular}{ll}
\toprule
\textbf{Participant} & \textbf{Flag States} \\
\midrule
PP1 & Bahamas, Liberia, Panama, Singapore\\
PP2 & Canada, United States\\
P1 & Canada\\
P2 & Panama, United States\\
P3 & Malta, Marshall Islands, United Kingdom\\
P4 & United States\\
P5 & Belgium, United States\\
P6 & Angola, Chile, Germany, Japan, Netherlands, New Zealand, \\
& South Africa, South Korea, United Kingdom, United States\\
P7 & United States\\
P8 & United States\\
P9 & Marshall Islands\\
P10 & United States\\
P11 & United States\\
P12 & United States\\
P13 & United States\\
P14 & Cook Islands, United States\\
P15 & Bahamas, Greece, Liberia, Norway, United States\\
P16 & United Kingdom, United States\\
P17 & Marshall Islands, United States\\
P18 & United States\\
P19 & Singapore\\
P20 & United States\\
P21 & United States\\
\bottomrule
\end{tabular}
\caption{Flags under which participants have worked/trained}
\label{tab:flag_by_participant}
*Participants are labeled as PP1 and PP2 for pilot participants, and P1–P21 for study participants.*
\end{table}

\end{document}